\documentclass[preprint,3p,sort&compress,a4paper]{elsarticle}

\usepackage{graphicx}
\usepackage{subfig}
\usepackage{enumerate} 
\usepackage{amssymb}
\usepackage{amsthm}

\usepackage[dvipdfm]{hyperref}
\hypersetup{
    colorlinks,%
    citecolor=black,%
    filecolor=black,%
    linkcolor=black,%
    urlcolor=black
}

\begin{document}

\begin{frontmatter}

\title{Verifying the self-affine nature of regional seismicity using nonextensive Tsallis statistics}

\author[1]{G. Minadakis}
\author[2]{S. M. Potirakis}
\author[1]{J. Stonham}
\author[3]{C. Nomicos}
\author[4]{K. Eftaxias}

\address[1]{Department of Electronic and Computer Engineering, Brunel University, Kingston Lane, Uxbridge, Middlesex, UB8 3PH, U.K.}
\address[2]{Department of Electronics, Technological Educational Institute of Piraeus, 250 Thivon \& P. Ralli, GR-12244, Aigaleo, Athens, Greece}
\address[3]{Department of Electronics, Technological Educational Institute of Athens, Ag. Spyridonos, Egaleo, GR 12210, Athens, Greece}
\address[4]{Department of Physics, Section of Solid State Physics, University of Athens, Panepistimiopolis, GR 15784, Zografos, Athens, Greece}

\begin{abstract}
The aspect of self-affine nature of faulting and fracture is widely documented from the data analysis of both field observations and laboratory experiments. In this direction, Huang and Turcotte have stated that the statistics of regional seismicity could be merely a macroscopic reflection of the physical processes in earthquake source, namely, the activation of a single fault is a reduced self-affine image of regional seismicity. This work verifies the aforementioned proposal. More precisely we show that the population of: (i) the earthquakes that precede of a significant event and occur around its the epicentre,  and (ii) the ''fracto-electromagnetic earthquakes'' that are emerged during the fracture of strong entities distributed along the activated single fault sustaining the system follow the same statistics, namely, the relative cumulative number of earthquakes against magnitude. The analysis is mainly performed by means of a recently introduced nonextensive model for earthquake dynamics which leads to a Gutenberg-Richter type law. We examine the variation of the parameters $q$ and $\alpha$, which are included in the nonextensive law, for different thresholds of magnitude and different radius around the epicentre. Such analysis enhances the physical background of the underlying self-affinity. The parameter $q$ describes the deviation of Tsallis entropy from the extensive Boltzmann-Gibbs entropy, and $\alpha$ is the constant of proportionality between the energy released during the fracture of a fragment and its size.
\end{abstract}


\end{frontmatter}

\section{Introduction}
\label{sec:intro}

The self-affine nature of fracture and faulting has been widely documented from the analysis from both laboratory and field observations \citep{Mandelbrot1982,Huang1988,Turcotte1997,Sornette2004,Muto2007}. Characteristically, Huang and Turcotte \citep{Huang1988} suggested that the statistics of regional seismicity could be merely a macroscopic reflection of the physical processes in the earthquake (EQ) source. This suggestion, from the perspective of self-affinity, implies that the activation of a single fault is a reduced self-affine image of regional seismicity. In this work we attempt to verify the aforementioned proposal drawing from a recently introduced model for EQ dynamics \citep{Sotolongo2004,Silva2006} which is based on the nonextensive Tsallis statistical approach \citep{Tsallis1988,Tsallis2001,Tsallis2009}. Its theoretical ingredient concerns two rough profiles interacting via the fragments filling the gap between them. The model leads to a nonextensive Gutenberg \& Richter (G-R) type formula that describes the frequency distribution of EQs against magnitude including two parameters: (i) the entropic index $q$, which describes the deviation of Tsallis entropy from the standard Boltzmann-Gibbs entropy, namely the index $q$ interprets the degree of non-extensivity, that accounts for the case of many non-independent, long-range interacting subsystems, and (ii) the physical quantity $\alpha $, which is a constant of proportionality between the energy released during the fracture of a fragment and its size $r$. 

The question naturally arises to whether the self-affine nature of fracture and faulting can be adequately explained by the nonextensive G-R type formula. In this direction we examine whether the nonextensive G-R type equation fit the following two populations of fracture events: (i) The fractures-EQs that precede a significant seismic event occurring in the region which surrounds its epicenter (foreshock activity). (ii) The fractures of strong entities which are distributed along the single activated significant fault sustaining the system. 

At this point the following question is emerged: \textit{``How can we know the sequence of magnitudes of fractures of strong entities which are evolved during the activation of a significant fault?''}. Crack propagation is the basic mechanism of material failure. It has been shown that during the mechanical loading, fracture induced acoustic and electromagnetic (EM) fields have been observed which allow the real-time monitoring of damage evolution in these materials. EM emissions in a wide frequency spectrum ranging from kHz to MHz are produced by opening cracks, which can be considered as the so-called precursors of general fracture. The radiated EM precursors are detectable both at a laboratory \citep{Bahat2005,Ogawa1985,Okeefe1995,Lolajicek1996,Panin2001,Frid2003,Mavromatou2004,Fukui2005,Lacidogna2011,Baddari2011} and geophysical scale \citep{Warwick1982,Gokhberg1982,Hayakawa1994,Hayakawa1999,Eftaxias2000,Eftaxias2001a,Hayakawa2002,Nagao2002,Eftaxias2002,Eftaxias2003,Eftaxias2004,Kapiris2004,
Contoyiannis2005,Karamanos2006,Eftaxias2007a,Papadimitriou2008,Kalimeri2008,Contoyiannis2008,Eftaxias2009,Eftaxias2010uni,Pulinets2003,Freund2007_PartII,Gokhberg1995,Morgounov2007}. 

An important feature, observed at both scales, is that {\it the MHz radiation precedes the kHz one} \citep{Eftaxias2002,Eftaxias2004,Kapiris2004,Contoyiannis2005,Eftaxias2009,Eftaxias2010}. Studies on the small (laboratory) scale reveal that the kHz EM emission is launched in the tail of pre-fracture EM emission from 97\% up to 100\% of the corresponding failure strength \citep[and references therein]{Eftaxias2002}. At the geophysical scale the kHz EM precursors are also emerged in the tail of preseismic EM emission, namely, from a few days up to a few hours before the EQ occurrence \citep{Kapiris2004}. Thus, the association of MHz, kHz EM precursors with the last stages of EQ generation seems to be justified. 

Based on the above mentioned experimental facts, a two-stage model has been recently proposed concerning the detected EM emissions in the field \citep{Kapiris2004,Contoyiannis2005,Contoyiannis2008,Papadimitriou2008,Eftaxias2006,Eftaxias2007b} suggesting that: 

\begin{enumerate}[(i)]
\item {The MHz EM emission is due to the fracture of the highly heterogeneous system that surrounds the fault. More specifically, the MHz EM activity can be attributed to phase transition of second order \citep{Contoyiannis2005}, while a Levy walk type mechanism can explain the observed critical state \citep{Contoyiannis2008}. We note that the heterogeneity and long-range correlations are two of the key components that make material failure an interesting field for the application of statistical mechanics. The suggestion that ruptures of heterogeneous systems is a critical phenomenon has been proposed by numerous of authors, e.g \citep{Allegre1982,Chelidze1982,Herrmann1990,Sornette1992,Lamaignere1996,Andersen1997,Sornette1990,Andersen1997,Sornette2004,Girard2012}
}

\item {The kHz EM phenomenon is rooted in the final stage of EQ generation, namely, the fracture of backbone of entities that prohibit the relative slipping of the two profiles of the fault \citep{Kapiris2004,Contoyiannis2005,Kalimeri2008,Papadimitriou2008,Eftaxias2009}. Thus, the detected precursory sequence of "fracto-EM kHz earthquakes" (see sec. \ref{sec:ath_em} ) represents the population of fractures that occur during the relative displacement of fault plates. 
}
\end{enumerate}

We note that the aforementioned proposal has been supported by a multidisciplinary analysis, e.g.,  in terms of extensive and non extensive statistical physics \citep{Karamanos2005,Papadimitriou2008,Kalimeri2008,Contoyiannis2008}, information theory, complexity \citep{Karamanos2006,Potirakis2011a}, laboratory experiments \citep{Eftaxias2002,Eftaxias2007a,Eftaxias2007b}, fault modeling \citep{Eftaxias2001a}, fractal electrodynamics \citep{Eftaxias2004}, self-affinity in fracture and faulting \citep{Eftaxias2010uni}, nonextensive model for earthquake dynamics \citep{Papadimitriou2008,Minadakis2011a,Minadakis2012a}, and mesomechanics \citep{Eftaxias2007b}. We note that the strong impulsive kHz EM time series shows strong persistent behaviour mirroring a non-equilibrium process without any footprint of an equilibrium thermal phase transition. We also note that fracture surfaces have been found to be self-affine following the persistent fractional Brownian motion (fBm) model over a wide range of length scales, while, the spatial roughness of fracture surfaces $H \simeq 0.7$ has been interpreted as a universal indicator of surface fracture, weakly dependent on the nature of the material and on the failure mode \citep[and references therein]{Lopez1998,Hansen2003,Ponson2006,Mourot2006,Zapperi2005}.

In this work, we attempt to provide evidence that the activation of a significant single fault is a reduced self-affine image of the regional seismicity, namely, the foreshock neighboring activity associated with the activation of the main fault, as follows: We first examine whether the aforementioned nonextensive G-R type formula can adequately describe both: 
(i) the populations of EQs included in different radius around the epicenter of a significant seismic event (foreshock activity), and (ii) the populations EM-EQs (see sec. \ref{sec:ath_em}) mirroring the fracture of strong entities distributed along the main fault that sustain the system. Furthermore, we focus on the variation of parameters $q$ and $\alpha$ included in the nonextensive formula for both populations using different thresholds of magnitudes. In this direction, two well documented cases that fulfil the needs of such analysis, with available data for both the seismicity and preseismic kHz EM emissions observed prior to large EQs, are: (i) the case of Athens EQ ($M=5.9$) occurred on 07-Sep-1999, in Greece and (ii) the case of L'Aquila EQ ($M = 6.3$) occurred on 06-Apr-2009 in Italy. 

This paper is organized as follows: In Sec. \ref{sec:theo}, we describe the basic principles of Tsallis nonextensive statistical mechanics and the nonextensive model for EQ dynamics. In Secs. \ref{sec:laquila} and \ref{sec:athens}, we apply nonextensive analysis on both the seismicity and preseismic kHz EM emissions, for the case of L'Aquila EQ and Athens EQ, respectively. In Sec. \ref{sec:analysis} we analyze our results providing further arguments that support the self-affine nature of fracture and faulting. Finally in Sec. \ref{sec:conclusions}, we summarize the key findings.

\section {Theoretical background}
\label{sec:theo}

The aim of statistical mechanics is to establish a direct link between the mechanical laws and classical thermodynamics. Within that context, ``extensivity'' is one of the crucial properties of the Boltzmann-Gibbs entropy ($S_{B-G}$) expressing the proportionality with the number of elements of the system. $S_{B-G}$ satisfies this rule, if the subsystems are statistically (quasi-) independent, or typically if the correlations within the system are essentially local. In such cases the system is called extensive. However, in the cases where the correlations may be far from negligible at all scales, the $S_{B-G}$ is ``nonextensive''. Inspired by multi-fractal concepts, Tsallis \citep{Tsallis1988} proposed a generalization of the B-G statistical mechanics, by introducing an entropic expression characterized by an index $q$ which leads to a nonextensive statistics: 
 
\begin{equation}
S_{q}=k\frac{1}{q-1}\left(1-\sum_{i=1}^{W}p_{i}^{q}\right),
\end{equation}

\noindent where $p_{i}$ are the probabilities associated with the microscopic configurations, $W$ is their total number, $q$ is a real number, and $k$ is Boltzmann's constant. $q \rightarrow 1$ corresponds to the standard extensive B-G statistics. Indeed, using $p_{i}^{(q-1)}=e^{(q-1)\ln(p_{i})}\sim 1+(q-1)\ln(p_{i})$ in the limit $q\rightarrow 1$, we obtain the standard B-G entropy:

\begin{equation}
S_{1}=-k\sum_{i=1}^{W}p_{i}\ln(p_{i}) \nonumber
\end{equation}

\noindent The entropic index $q$ characterizes the degree of non-additivity reflected in the following pseudo-additivity rule: $S_{q}(A+B)=S_{q}(A)+S_{q}(B)+(1-q)S_{q}(A)S_{q}(B)$. 


\noindent The cases $q>1$ and $q<1$, correspond to sub-additivity, or super-additivity, respectively. The parameter $q$ itself is not a measure of the complexity of a time series. It measures the degree of nonextensivity of the corresponding system. A metric of the dynamic changes of the complexity of a system is the time variations of the Tsallis entropy for a given $q$ ($S_{q}$).

\subsection{A fragment-asperity model for earthquakes coming from a nonextensive Tsallis formulation}

The best known scaling relation for EQs is the Gutenberg \& Richter (G-R) magnitude-frequency relationship \citep{Gutenberg1954}, given by: 

\begin{equation}\label{eq:b-value}
\log N(>m)=\alpha-bm,
\end{equation}

\noindent where $N(>m)$ is the cumulative number of EQs with a magnitude greater than $m$ occurring in a specified area and time. Parameters $b$ and $\alpha$ are constants.  

A nonextensive model for EQ dynamics consisting of two rough profiles interacting via fragments filling the gap has been recently introduced by Sotolongo-Costa and Posadas (SCP) \citep{Sotolongo2004}. More recently, the aforementioned model was revised by Silva et al. \citep{Silva2006}, where two crucial ingredients were employed. They use revised definition for the mean values in the context of Tsallis nonextensive statistics that was achieved in the study of Abe and Bagci \citep{Abe2005}. Moreover, Silva et al. proposed  a new scaling law, $\varepsilon \propto r^3$, between the released relative energy $\varepsilon$ and the size $r$ of fragments. Finally, their approach leads to the following G-R type law for the magnitude distribution of EQs:

\begin{eqnarray}
G(>M) = \frac{N\left(>M\right)}{N} = \left({2-q\over 1-q} \right)\log \left[1-\left({1-q\over 2-q} \right)\left({10^{2M} \over a^{2/3} } \right)\right] \label{eq:silva}
\end{eqnarray}

\noindent where, $N$ is the total number of EQs, $N(>M)$ the number of EQs with magnitude larger than $M$, and $M\approx \log (\varepsilon)$. The parameter $\alpha$ is the constant of proportionality between the energy, $\varepsilon$, and the size of fragment, $r$. This is not a trivial result, and incorporates the characteristics of nonextensivity into the relative cumulative number of EQs against magnitude. The entropic index $q$ describes the deviation of Tsallis entropy from the traditional $B-G$ one, expressing the long-range correlations developed in the system. 

The $q$-parameter included in the non-extensive formula (Eq. (\ref{eq:silva})) is associated with the $b$ parameter of Gutenberg \& Richter formula (Eq. (\ref{eq:b-value})), by the relation \citep{Sarlis2010}:

\begin{equation} 
b=2\times \left( {2-q\over q-1}\right) 
\label{eq:sarlis}
\end{equation}

Note that the nonextensive model of Eq. (\ref{eq:silva}) has been successfully applied for several different regions under study incorporating the characteristics of nonextensivity statistics into the detected EQs \citep{Silva2006,Matcharashvili2011,Telesca2010,Telesca2010b,Telesca2010c,Telesca2011}. Values of $q \sim 1.6 - 1.8$ \citep{Silva2006,Vilar2007,Matcharashvili2011,Telesca2010,Telesca2010b} seem to be universal, in the sense that different data sets from different regions of the globe indicate a value for the nonextensive parameter lying in this interval. In the following sections the two aforementioned cases of strong EQs are examined, for both the seismicity and the observed preseismic kHz EM emissions.


\section{The case of L'Aquila 2009 earthquake}
\label{sec:laquila}

We recall that the prospect in this work is to examine whether the self-affine nature of fracture and faulting can be adequately explained by the nonextensive Eq. (\ref{eq:silva}). Eq. (\ref{eq:silva}) is directly connected to the traditional G-R law (Eq. \ref{eq:b-value}), above some magnitude threshold through Eq. (\ref{eq:sarlis}), which in turn leads to the power-low distribution of magnitudes, expressing the fractal nature of the system under study \citep{Sator2010,Turcotte1986,Sammis1986,Kaminski1998,Carpinteri2005}. Thus in the prospect to verify the aforementioned suggestion, in the following sections we examine both the seismicity (foreshock activity) and the observed preseismic kHz EM emissions related to large EQs, in the prospect to verify the aforementioned suggestion.

\subsection{Analysis of seismicity (L'Aquila case)}
\label{sec:lac_seis}
The analysis here is focused on the very shallow strong (Mw = 6.3) earthquake occurred on 6 April 2009 in L'Aquila, Central Italy, at 01:32:39 UTC. Characteristically, Papadopoulos et al. \citep{Papadopoulos2010}, reported that from the beginning of 2006 up to the end of October 2008 no particular earthquake activation was noted in that seismogenic area. On the contrary, from 28-October-2008 up to 27-March-2009 the seismicity was in the state of weak foreshock activity, and dramatically increased 10 days prior the main event (from $\approx$ 26-Mar-2009) \citep{Papadopoulos2010}. Drawing from these evidence, the period from 28-Oct-2008 00:00:00 up to 6-April-2009 01:32:00 was selected as the most appropriate time-frame in order to examine the self-affine behaviour of the EQs included within different radii around the EQ epicenter. The Italian EQ catalogue which is available on the website of the \textit{Istituto Nazionale di Geofisica e Vulcanologia} (INGV: \url{http://bollettinosismico.rm.ingv.it}), was used. 

\begin{figure}[h]
	\centering
	\subfloat[0-400km]{\label{subfig:q_lac1}\includegraphics[width=0.33\textwidth]{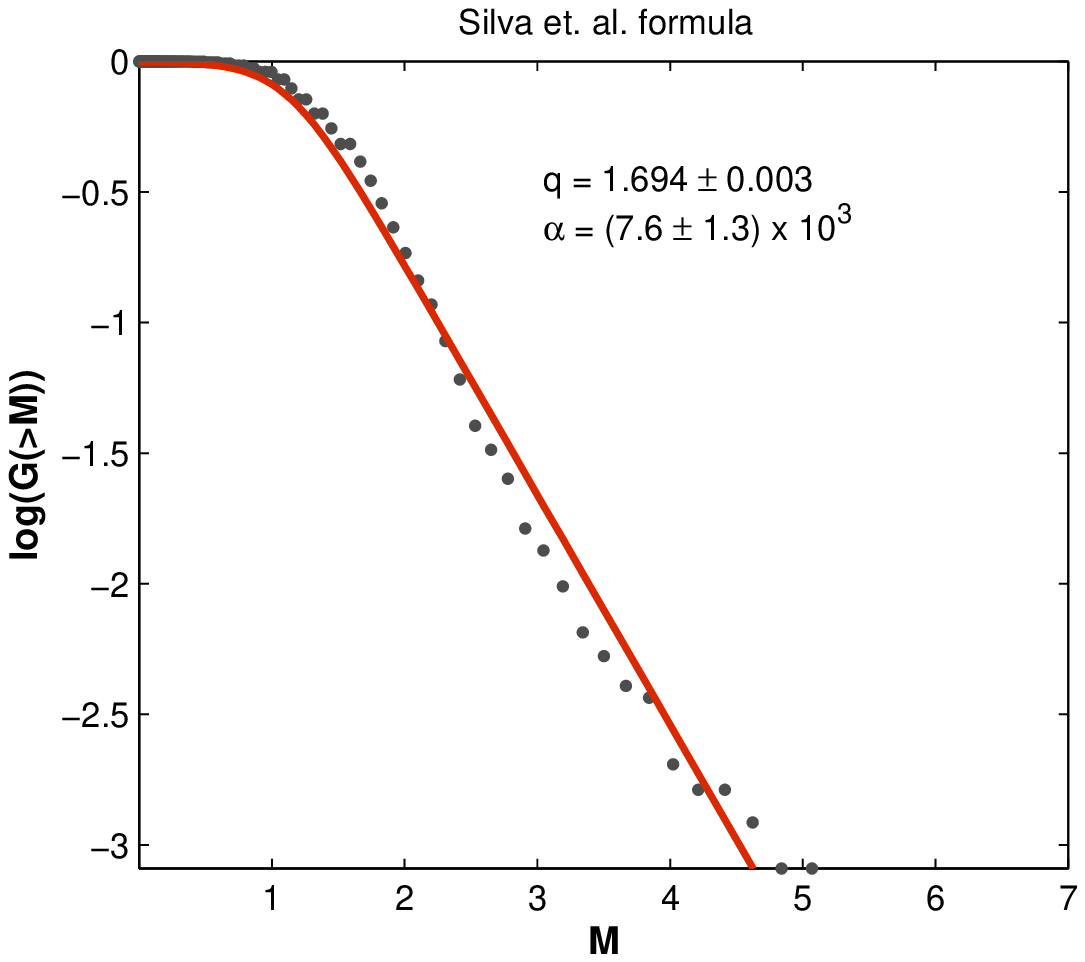}}             
	\subfloat[0-300km]{\label{subfig:q_lac2}\includegraphics[width=0.33\textwidth]{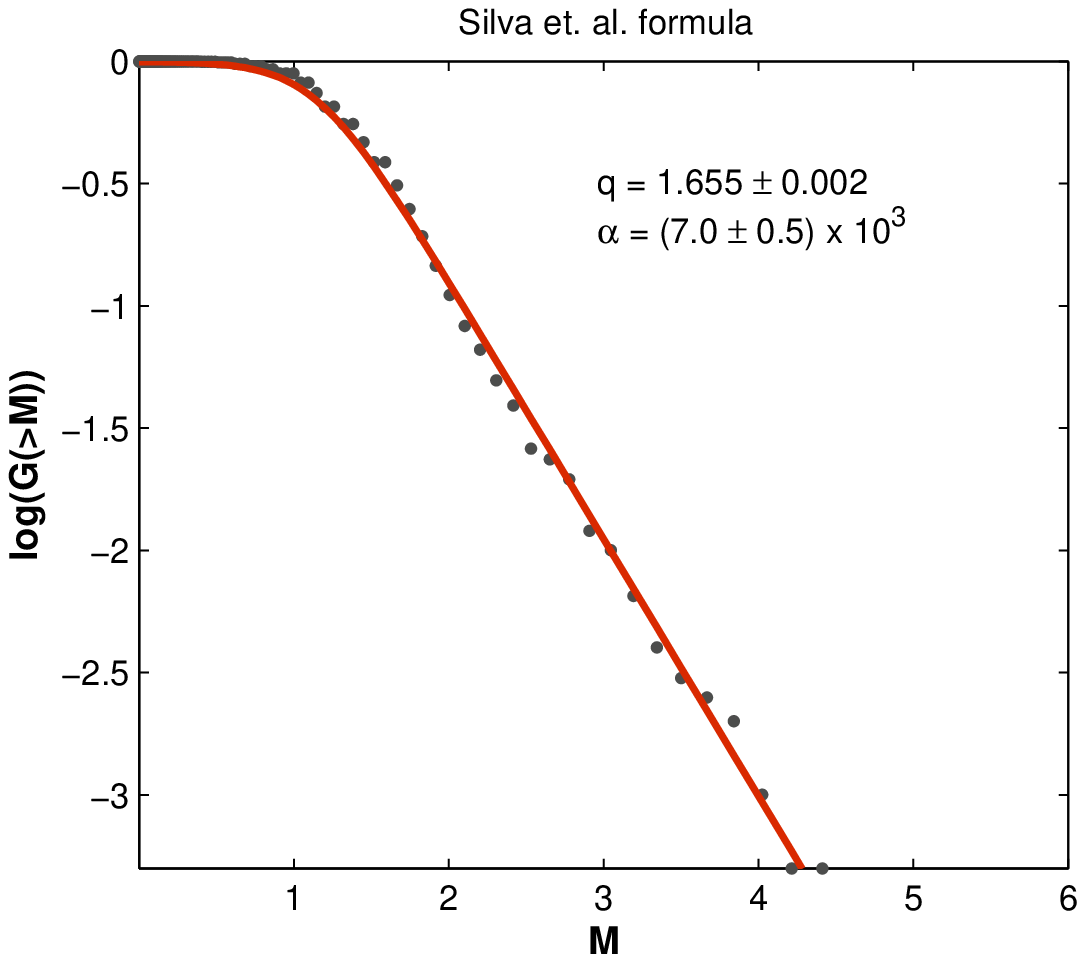}}      
	\subfloat[0-200km]{\label{subfig:q_lac3}\includegraphics[width=0.33\textwidth]{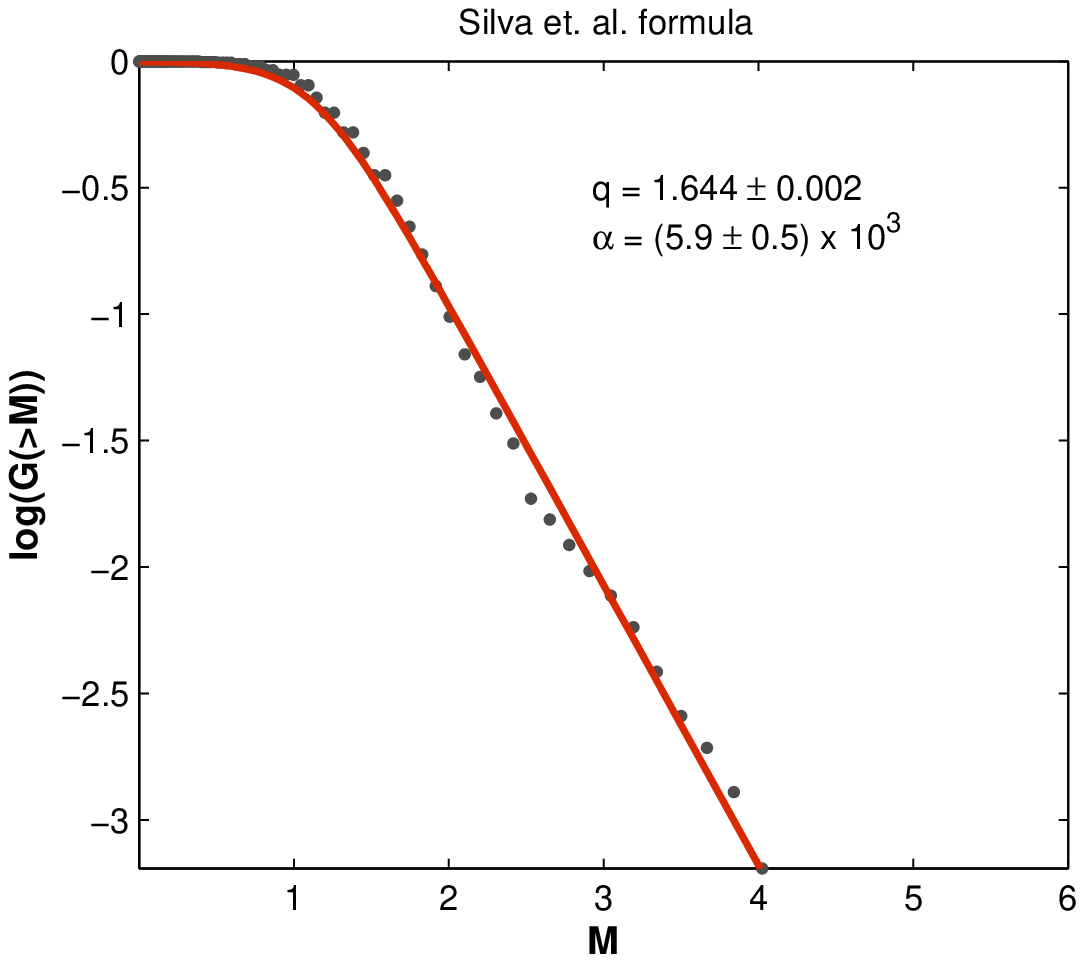}}      
       
	\subfloat[0-100km]{\label{subfig:q_lac4}\includegraphics[width=0.33\textwidth]{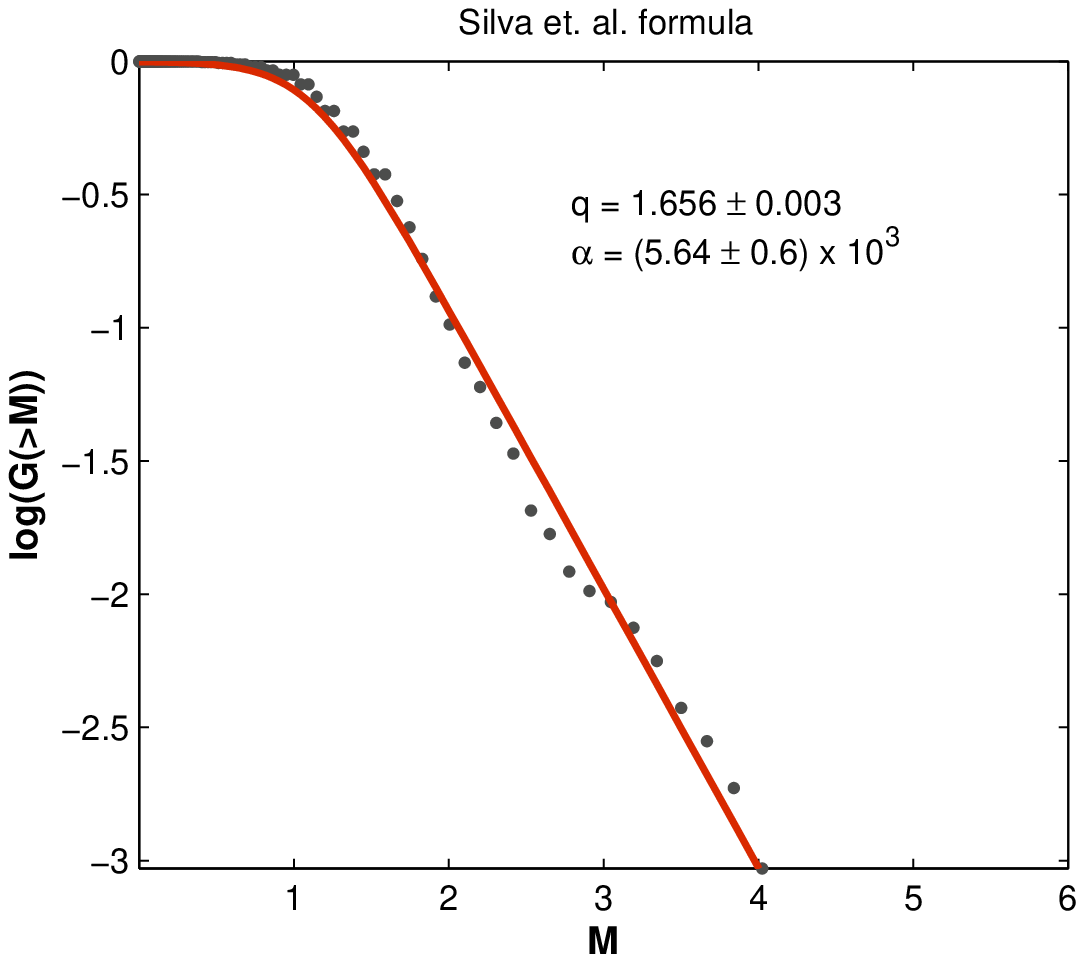}}             
	\subfloat[0-50km]{\label{subfig:q_lac5}\includegraphics[width=0.33\textwidth]{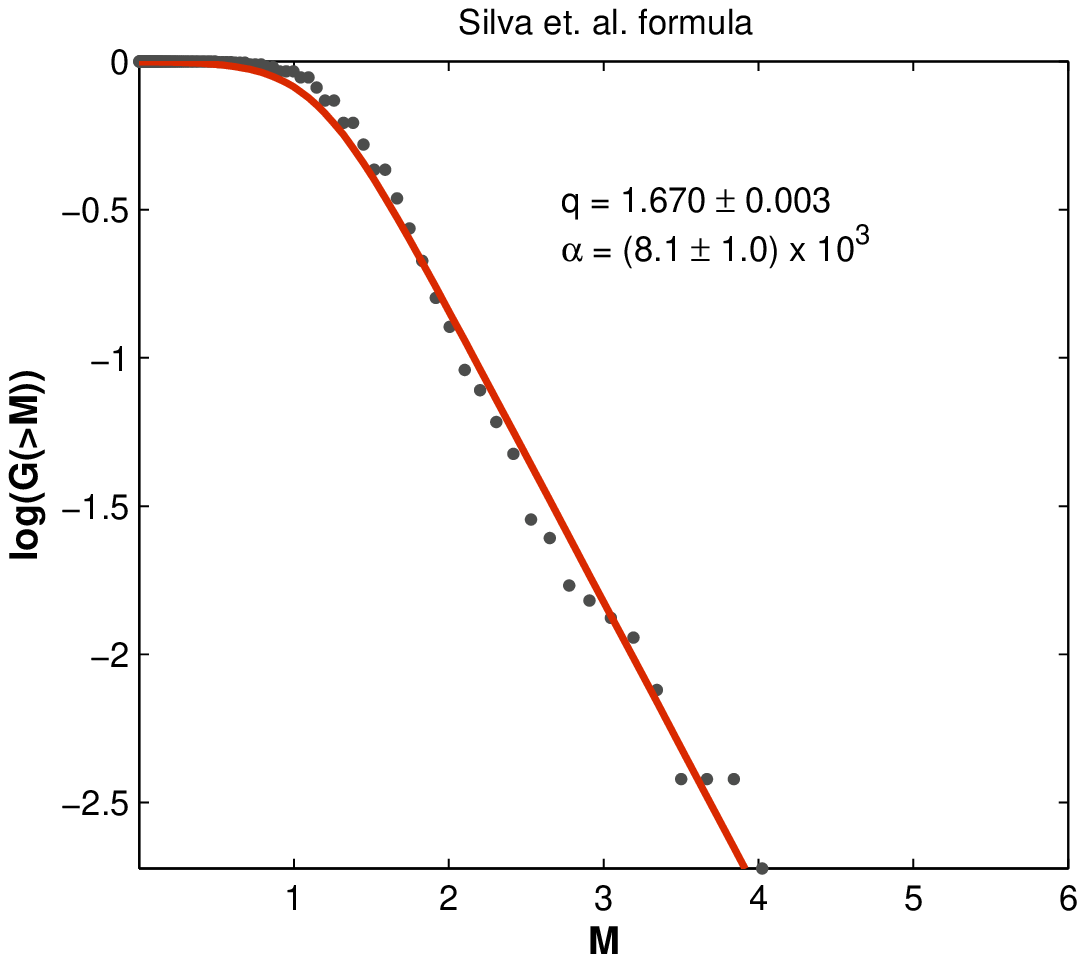}}      
	\subfloat[0-30km]{\label{subfig:q_lac6}\includegraphics[width=0.33\textwidth]{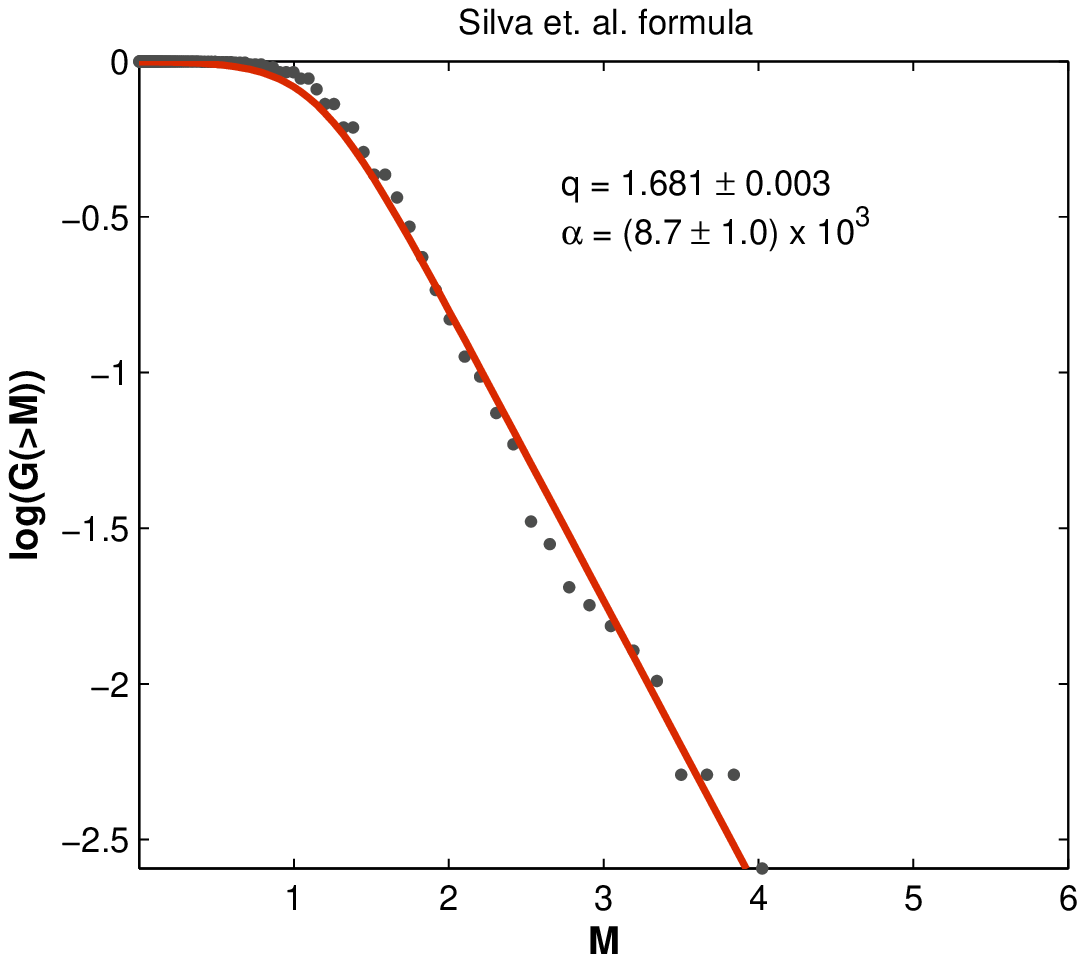}}      

	\caption{We use Eq. (\ref{eq:silva}) to fit the foreshock seismic data in terms of the relative cumulative number of EQs included in six different geographic areas around the L'Aquila EQ epicenter.}
\label{fig:q_lacuila}
\end{figure}

Herein, we first examine whether the aforementioned nonextensive G-R type formula (Eq. (\ref{eq:silva})) can adequately describe the populations of EQs included within different radii around the epicenter of L'Aquila EQ. In this direction Eq. (\ref{eq:silva}) was used to fit the seismic data of six different selected geographic areas around the EQ epicenter, for the period under study: 0-400km, 0-300km, 0-200km, 0-100, 0-50km and 0-30km respectively, as shown in Fig. \ref{fig:q_lacuila}. The Levenberg-Marquardt (LM) fitting method \citep{Levenberg1944,Marquardt1963} was used for optimal fitting. From Fig. \ref{fig:q_lacuila}, we observe that Eq. (\ref{eq:silva}) provides an excellent fit on the relative cumulative number of magnitudes ($G(>M)$) related to the populations of EQs included in each one of the six selected geographic areas. The nonextensive $q$-parameter varies between $[1.644 \sim 1.694]$ with a relative small standard error ranging between $[0.002 \sim 0.003]$. 

\begin{figure}[h]
	\centering
	\subfloat[0-400km]{\label{subfig:qa_lac1}\includegraphics[width=0.33\textwidth]{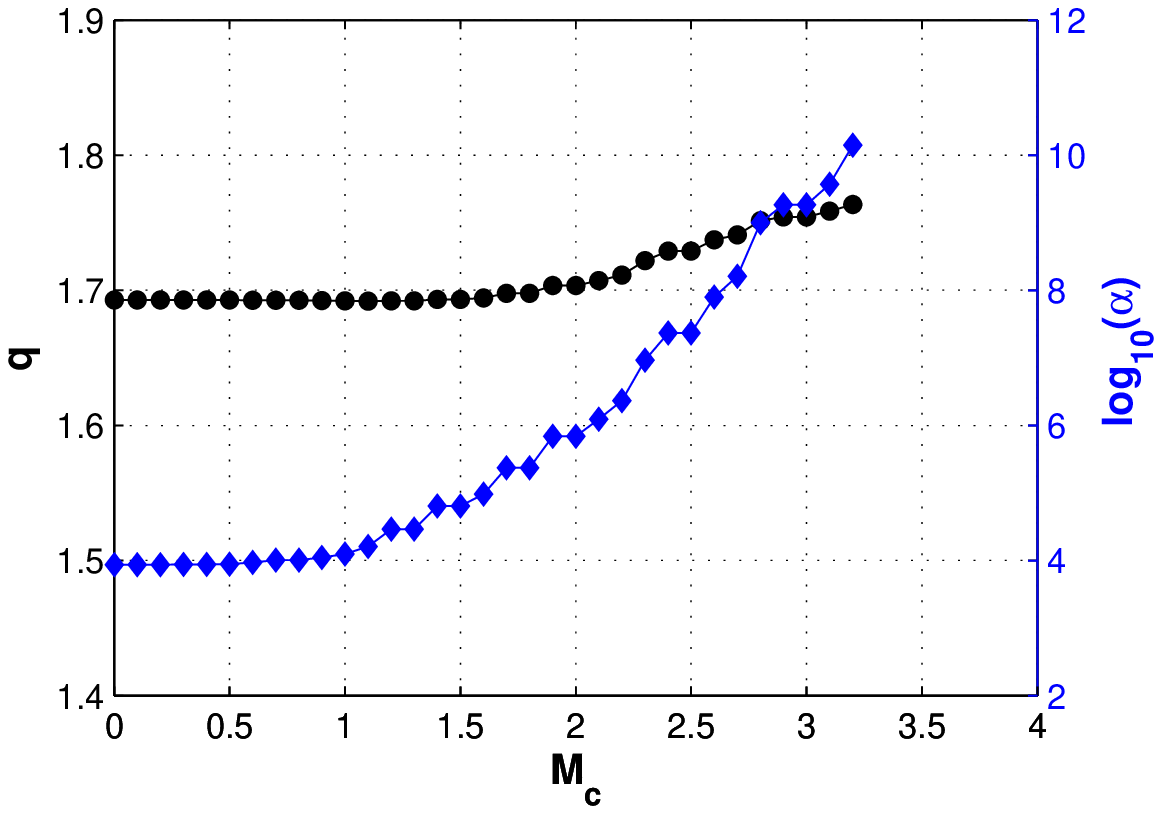}}             
	\subfloat[0-300km]{\label{subfig:qa_lac2}\includegraphics[width=0.33\textwidth]{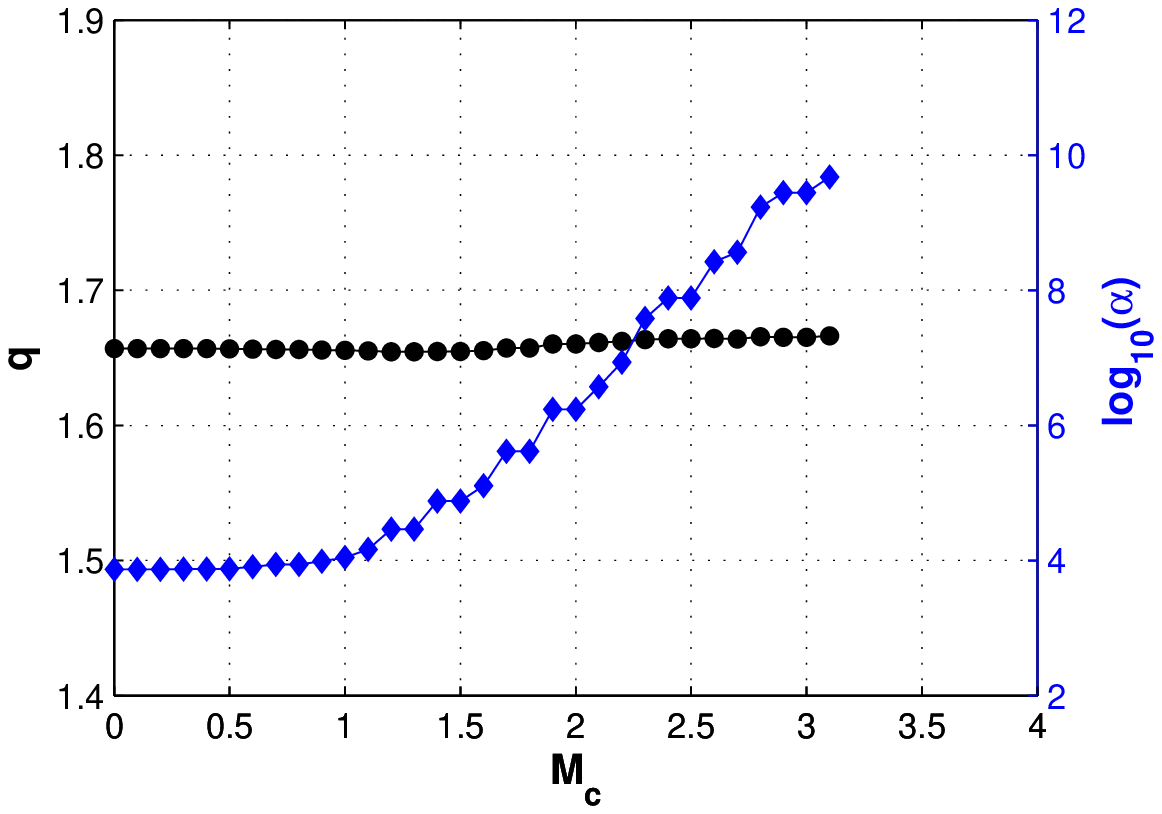}}      
	\subfloat[0-200km]{\label{subfig:qa_lac3}\includegraphics[width=0.33\textwidth]{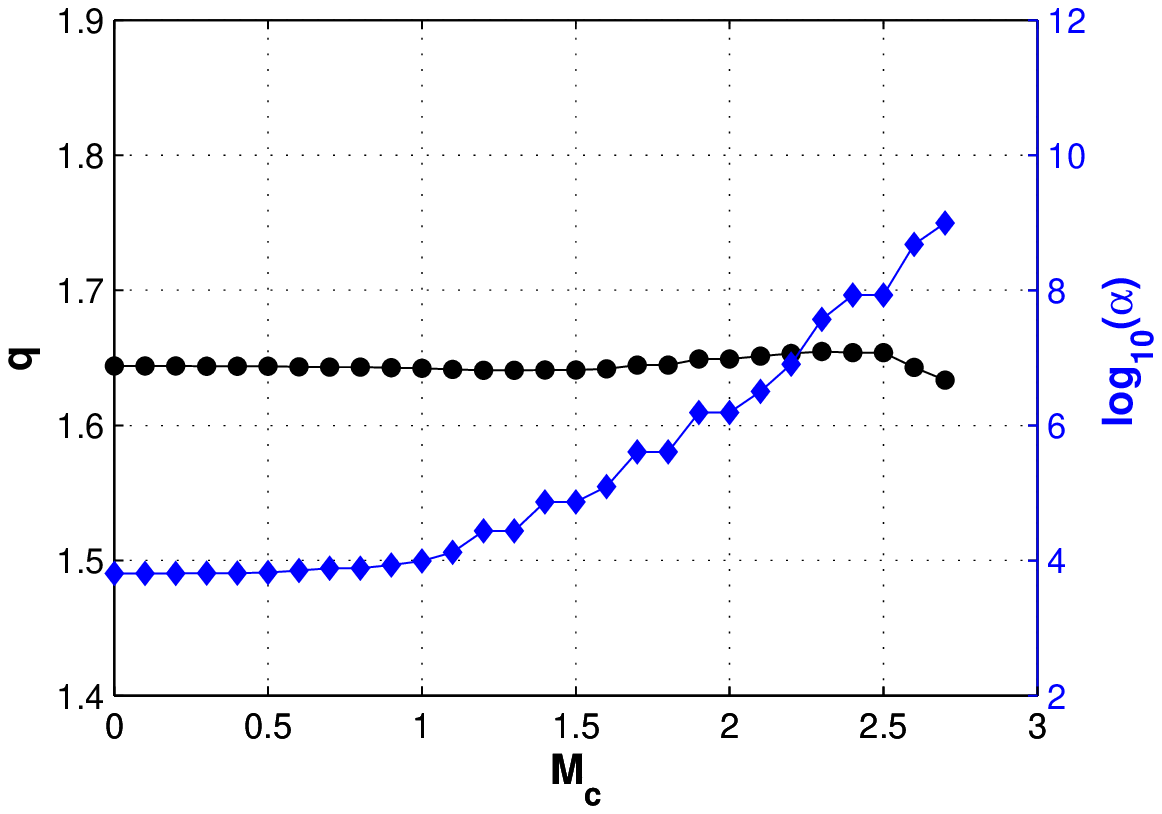}}      
       
	\subfloat[0-100km]{\label{subfig:qa_lac4}\includegraphics[width=0.33\textwidth]{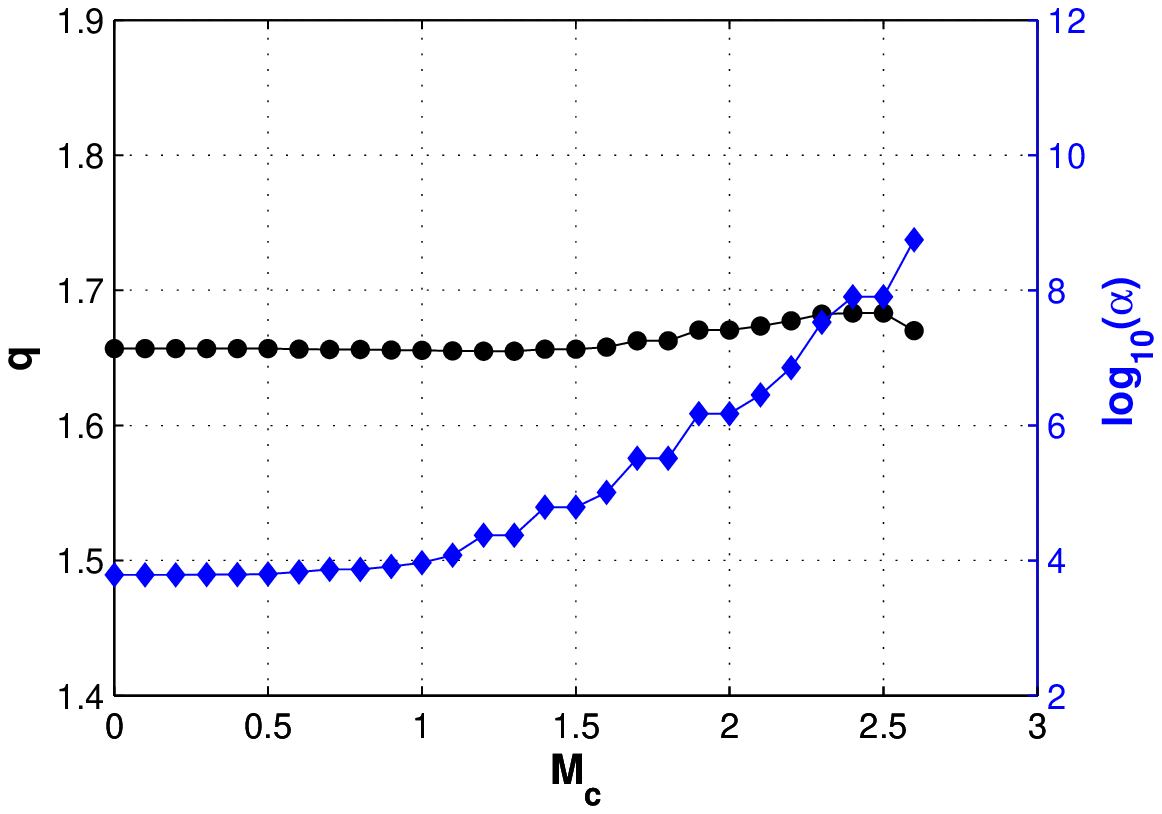}}             
	\subfloat[0-50km]{\label{subfig:qa_lac5}\includegraphics[width=0.33\textwidth]{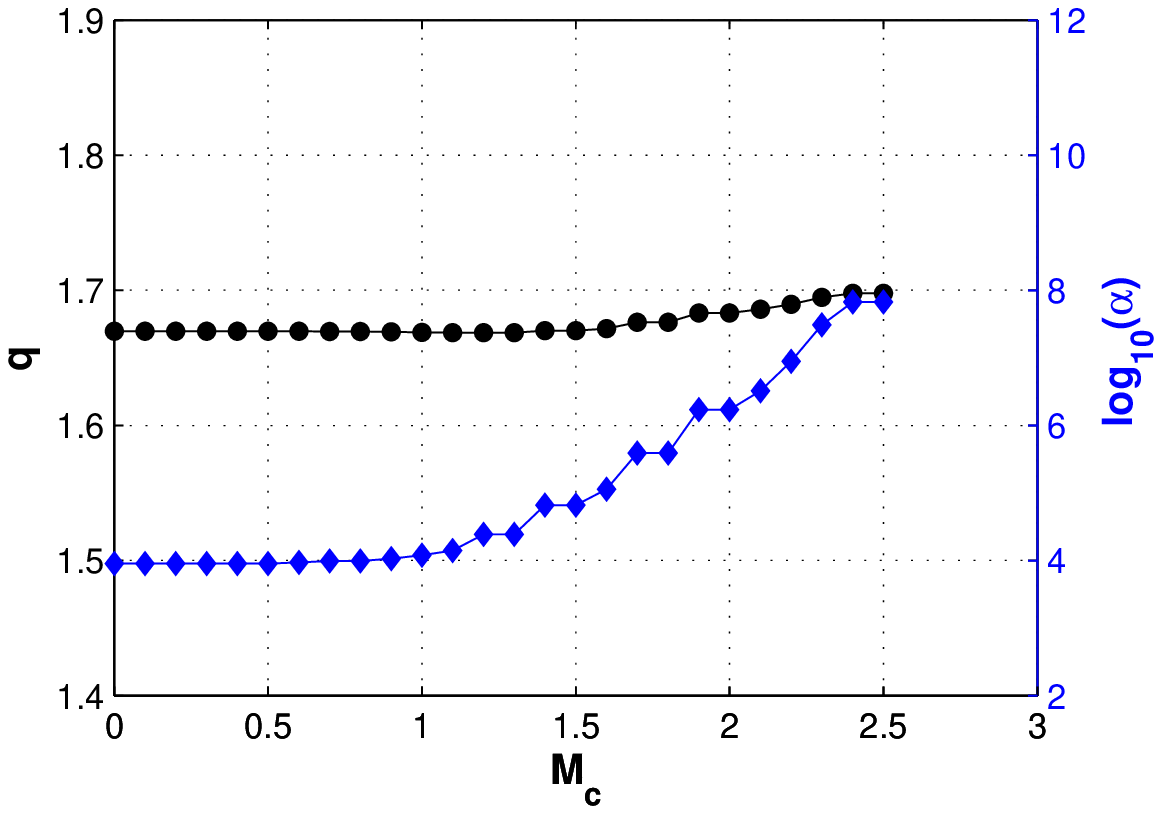}}      
	\subfloat[0-30km]{\label{subfig:qa_lac6}\includegraphics[width=0.33\textwidth]{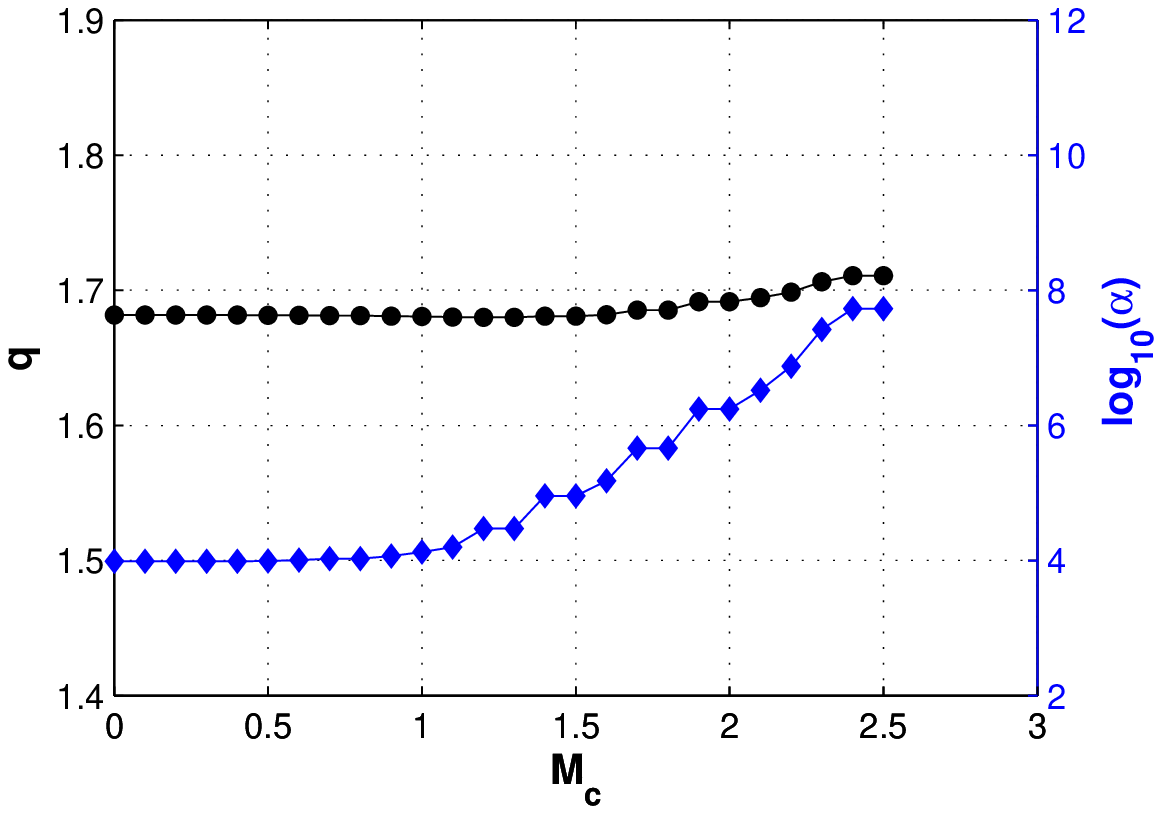}}      

\caption{(a) Variation of nonextensive parameter $q$ (see black-bullets) and the volumetric energy density $\alpha $ (see blue-rhombuses), for different thresholds of magnitudes of the detected EQs included in the period from 28-Oct-2008 00:00:00 up to 6-April-2009 01:32:00, for different geographic areas around the EQ epicenter: (a) 0-400km (b) 300km, (c) 0-200km (d) 0-100km (e) 0-50 and (f) 0-30km around the epicenter of L'Aquila EQ.}
\label{fig:qa_lacuila}
\end{figure}

Secondly, in the prospect to examine the behaviour of parameters $q$ and $\alpha$ included in the nonextensive formula, different thresholds of magnitudes cut-offs ($M_c$) were applied using an increasing step of $0.1$. For each step, formula (\ref{eq:silva}) along with the use of Levenberg-Marquardt (LM) fitting method, were used to fit the seismic data in terms of the relative cumulative number of EQs. A minimum number of 20 events was considered as a criterion for the statistical completeness of each calculation. The derived parameters $q$ and $\alpha $, were graphically placed on to a common $x$-axis chart as shown in Fig. \ref{fig:qa_lacuila}. It is observed that both the nonextensive $q$-parameter and the energy $\alpha$, present similar behaviour, for all the selected geographic areas around the EQ epicenter. More specifically, the nonextensive parameter $q$ (depicted with black bullets) remains relative stable with minor increment at intermediate thresholds of magnitudes ($M_c\approx 1.5 - 2.5$). The characteristic value of the volumetric energy density $\alpha $ (depicted with blue-rhombuses), increases at higher magnitude thresholds ($M_c$). 

\clearpage

As we have seen so far, the analysis of seismicity at different geographic scales revealed similar footprints for both the parameters included in the nonextensive formula. The same behaviour is expected to be found at lower scales of fracturing, namely, the preseismic kHz EM anomalies observed prior to L'Aquila EQ. Thus, in the following subsection, we examine these kHz EM emissions in terms of the nonextensive model.

\subsection{Analysis of preseismic kHz EM emissions observed prior to L'Aquila EQ}
\label{sec:lac_em}
This subsection focuses on the kHz EM emissions detected from a measurement station which has been installed and operating at a mountainous site of Zante Island in the Ionian Sea (Western Greece). During the aforementioned 10-day foreshock activity, well documented \citep{Eftaxias2009,Eftaxias2010}, preseismic kHz EM anomalies were observed on 4-Apr-2009, two days after the MHz EM anomalies of $2^{th}$ Apr, verifying the two-stage model described in Sec. \ref{sec:intro}. More precisely, the detected anomalies followed the temporal scheme listed below: 

\begin{enumerate}[(i)]
\item {The MHz EM anomalies were detected on 26 March 2009 and 2 April 2009.}
\item {The kHz EM anomalies emerged on 4 April 2009.}
\item {The ULF EM anomaly was continuously recorded from 29 March 2009 up to 2 April 2009.}
\end{enumerate}

\begin{figure}[h]
\begin{center}             
\subfloat[10 kHz NS]{\includegraphics[width=0.33\textwidth]{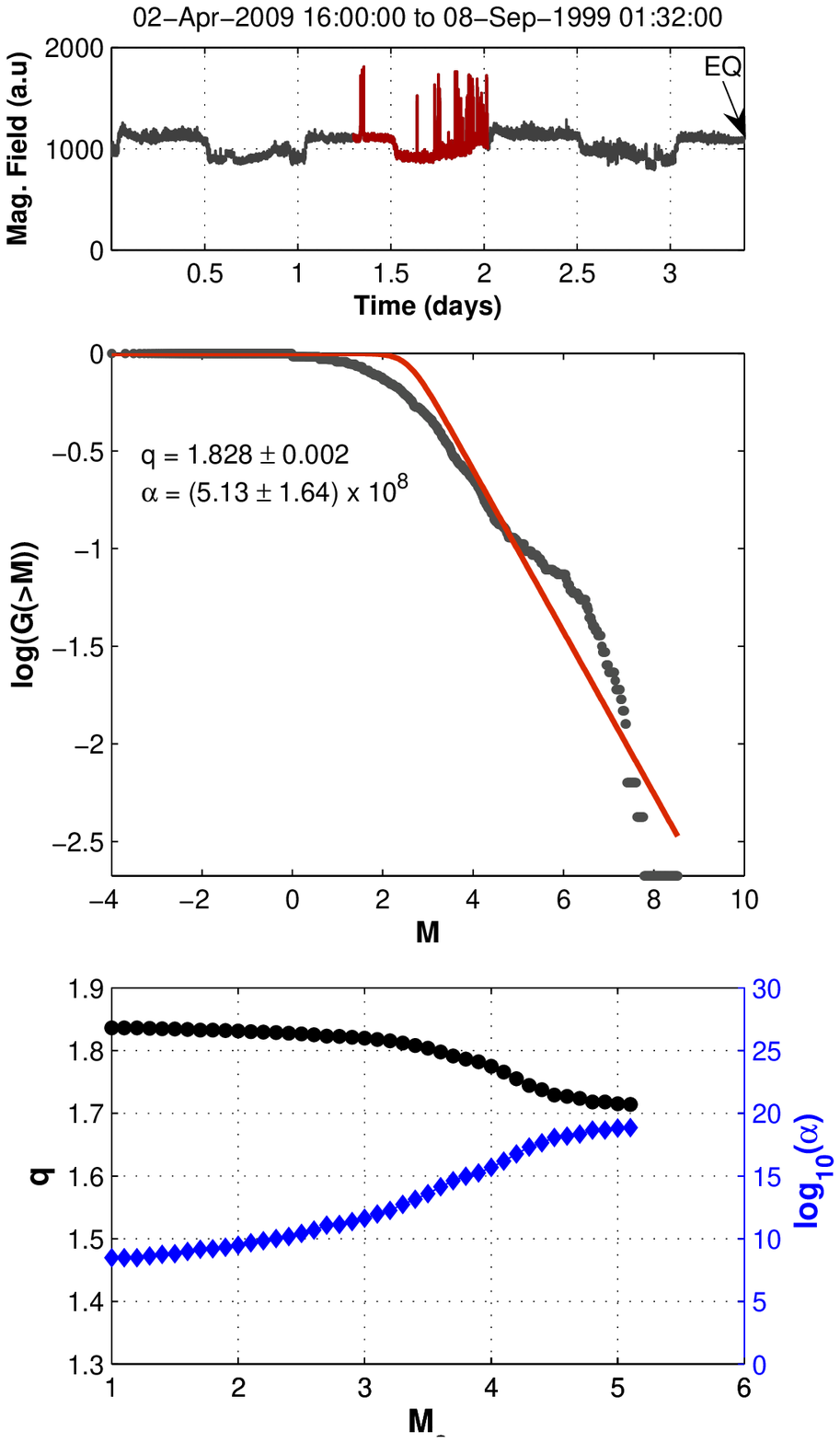}}
\subfloat[10 kHZ EW]{\includegraphics[width=0.33\textwidth]{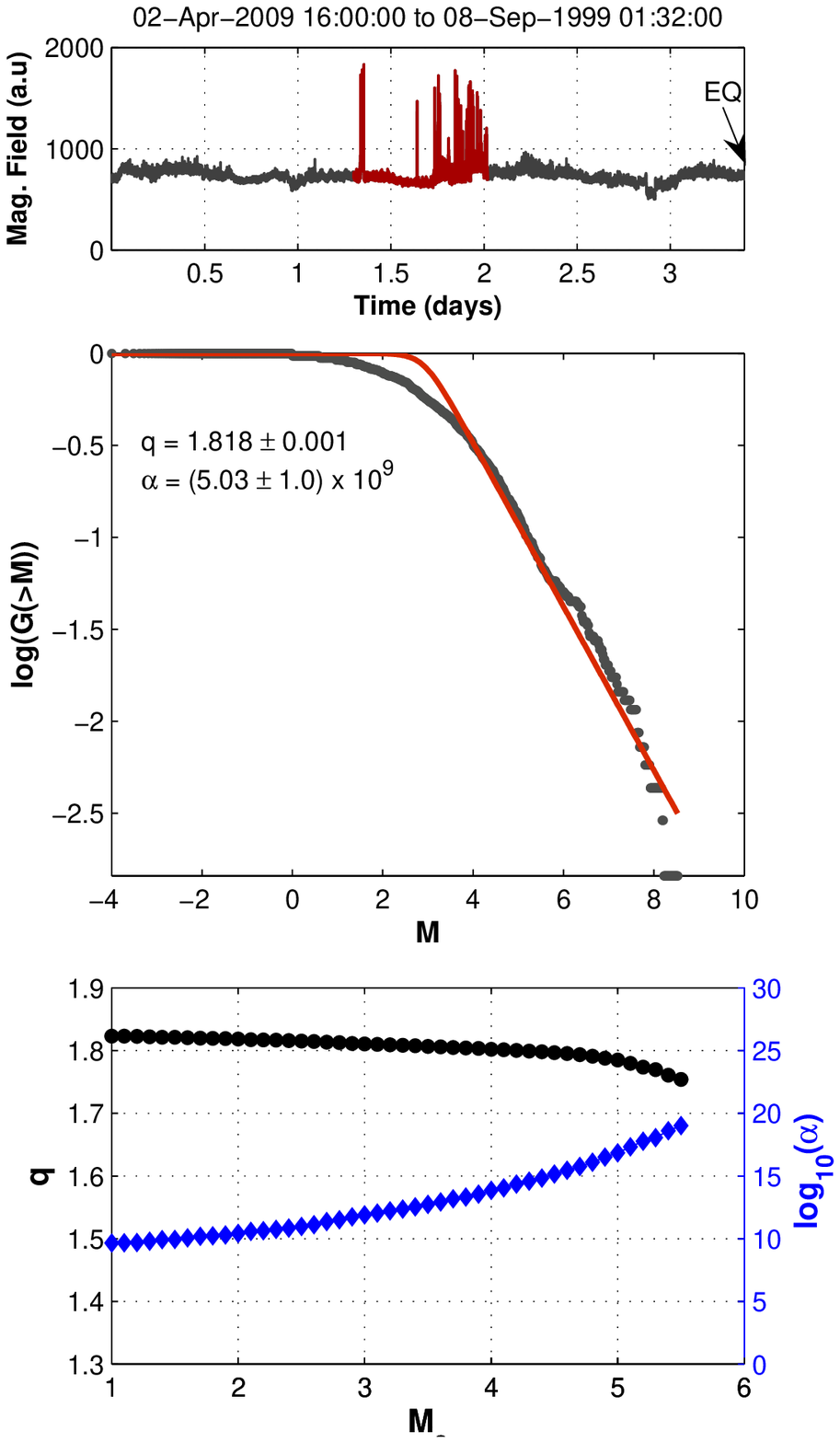}}
\subfloat[10 kHz V]{\includegraphics[width=0.33\textwidth]{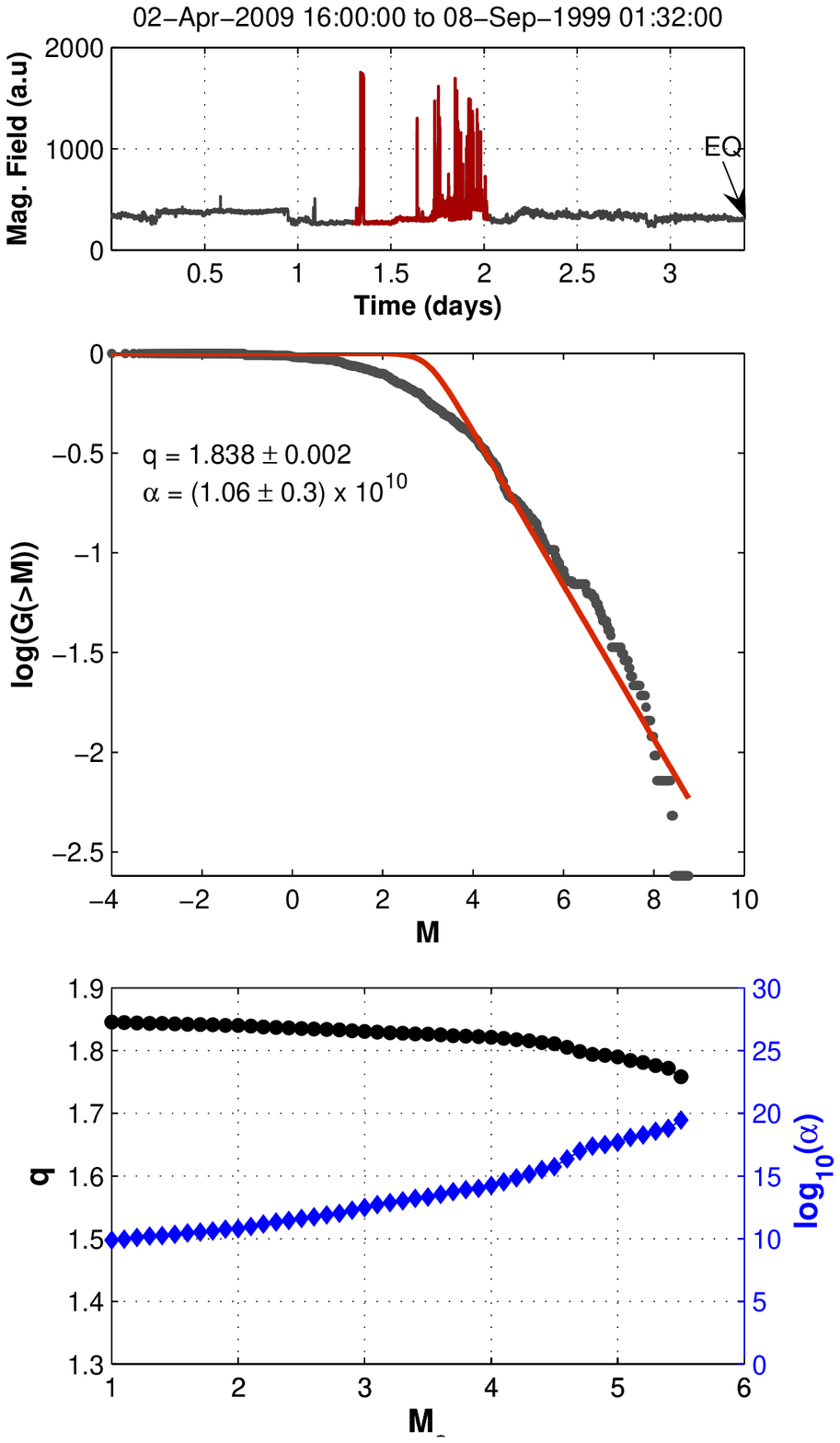}}

\end{center}
\caption{Three observed magnetic fields recorded on 02-Apr-2009 16:00:00 to 06-Apr-2009 01:32:39. The selected red parts of each recorded signal refer to the period where the kHz EM anomalies observed and have been well justified for their seismogenic origin (\citep{Kapiris2004,Contoyiannis2005,Contoyiannis2008,Papadimitriou2008}). The middle charts of each subfigure, refer to the fitting of the EM data in terms of the relative cumulative number of EM-EQs $G(> M)$ (see Eq. (\ref{eq:silva})), included in the period under study (red-part). The bottom charts of each sub-figure, refer to the corresponding variation of the nonextensive $q$-parameter (black-bullets) and the volumetric energy density $\alpha $ (blue-rohmbs), for different thresholds of magnitude cut-off ($M_c$).}
\label{fig:em_vars_lac}
\end{figure}

In Fig. \ref{fig:em_vars_lac}, the top charts of each sub-figure refer to the three observed magnetic field strengths, recorded from the 10 kHz (NS,EW,V) sensors, on 4-Apr-2009. Note that the EQ occurred on 06-Apr-2009 01:32:39 UTC as shown by the black-arrow. The selected red parts refer to the period where the kHz EM anomalies observed, which in turn have been well justified for their seismogenic origin from recent studies \citep{Eftaxias2009,Eftaxias2010}.

\textit{The notion of electromagnetic earthquake:} We regard as amplitude $A$ of a candidate ``fracto-electromagnetic fluctuation'' the difference $A_{fem}(t_{i})=A(t_{i})-A_{noise}$, where $A_{noise}$ is the background (noise) level of the EM time series. We consider that a sequence of $k$ successively emerged ``fracto-electromagnetic fluctuations'' $A_{fem}(t_{i})$, $i=1,\ldots,k$ represents the EM energy released, $\varepsilon$, during the damage of a fragment. We shall refer to this as an ``electromagnetic earthquake'' (EM-EQ). Since the squared amplitude of the fracto-electromagnetic emissions is proportional to their power, the magnitude $m$ of the candidate EM-EQ is given by the relation: 

\begin{equation}
 m=\log\varepsilon \sim \log
\left(\sum\left [ A_{fem}(t_{i})\right]^{2}\right)
\end{equation}

Herein, as in the case of seismicity, we first examine whether the aforementioned nonextensive G-R type formula (Eq. (\ref{eq:silva})) can adequately describe the populations of EM-EQs included in the preseismic kHz EM time series. As shown in the middle charts of Fig. \ref{fig:em_vars_lac}, we used Eq. (\ref{eq:silva}) to fit the EM data in terms of the relative cumulative number of electromagnetic earthquakes  $G(> M)$, included in the period that refers to the red part of these signals. It is observed that Eq. (\ref{eq:silva}) provides a satisfactory fit to the preseismic kHz EM experimental data associated with the L'Aquila EQ for all the three recorded components. The nonextensive $q$-parameter varies between $[1.818 \sim 1.838]$ with a relative small standard error ranging between $[0.001 \sim 0.002]$. 

Secondly, we apply Eq. \ref{eq:silva}, on the relative cumulative number of EM-EQ contained in the preseismic kHz EM emissions in the prospect to examine whether the variation of nonextensive $q$-parameter the volumetric density $\alpha $, also present the same behaviour with that of seismicity (see Sec. \ref{sec:lac_seis}). Different thresholds of magnitude cut-offs ($M_c$) were applied on the EM-EQs contained in each one of the aforementioned channels using an increasing step of $0.1$. For each step, Eq. (\ref{eq:silva}) along with the use of LM method, was used to fit the EM data in terms of the relative cumulative number of EM-EQs contained in each channel. In Fig. \ref{fig:em_vars_lac}, the bottom charts of each sub-figure, depict the variation of the nonextensive $q$-parameter (black bullets) and the volumetric energy density $\alpha $ (blue rhombuses), for different thresholds of magnitudes ($M_c$). It is observed that the variation of the nonextensive $q$-parameter remains relative constant with minor decrement at higher thresholds of magnitudes. On the contrary, the energy density $\alpha $ mirrors the behaviour of $q$-parameter, presenting a relative increment at higher thresholds. 

These latter results evidently show the similarity with the results obtained in Sec. \ref{sec:lac_seis}, indicating the self-affine nature of fracture and faulting, from the large scale of foreshock seismicity, to the fault-generation of a single EQ in terms of preseismic kHz EM emissions. However, in order to ensure the consistency of the results, in the following sections, we further examine the case of Athens 1999 EQ.

\clearpage
\section{Additional evidence of the self-affinity of EQs: The case of Athens 1999 earthquake}
\label{sec:athens}

\subsection{Analysis of seismicity (Athens case)}
\label{sec:ath_seis}
We focus on the case of Athens EQ ($Mw=5.9$) occurred on 07-Sep-1999 11:56:50 UTC, namely, the period from 17-Aug-1999 00:01:39.80 up to 07-Sep-1999 01:56:49. The Greek EQ catalog was used, as provided by the website of the Institute of Geodynamics of the National Observatory of Athens (\url{www.gein.noa.gr}). Note that Athens EQ occurred very shortly after the major 17/8/1999, $Mw = 7.4$, EQ which took place on Izmit, Turkey, approximately 650 km North-East of Athens. It has been shown that the Izmit EQ, was followed immediately by small earthquakes occurred throughout much of continental Greece \citep{Brodsky2000}. This explains the selected period. 

\begin{figure}[h]
	\centering
	\subfloat[0-400km]{\label{subfig:q_ath1}\includegraphics[width=0.4\textwidth]{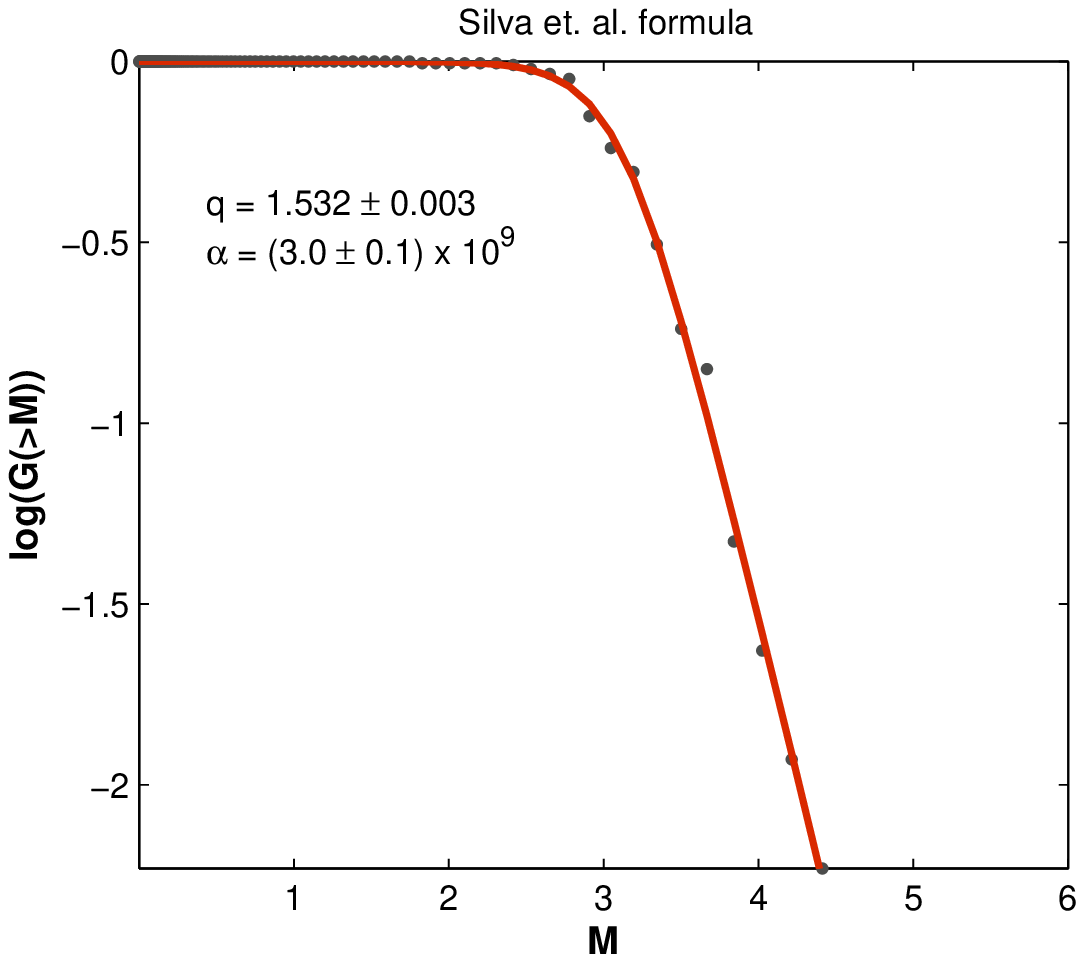}}             
	\subfloat[0-300km]{\label{subfig:q_ath2}\includegraphics[width=0.4\textwidth]{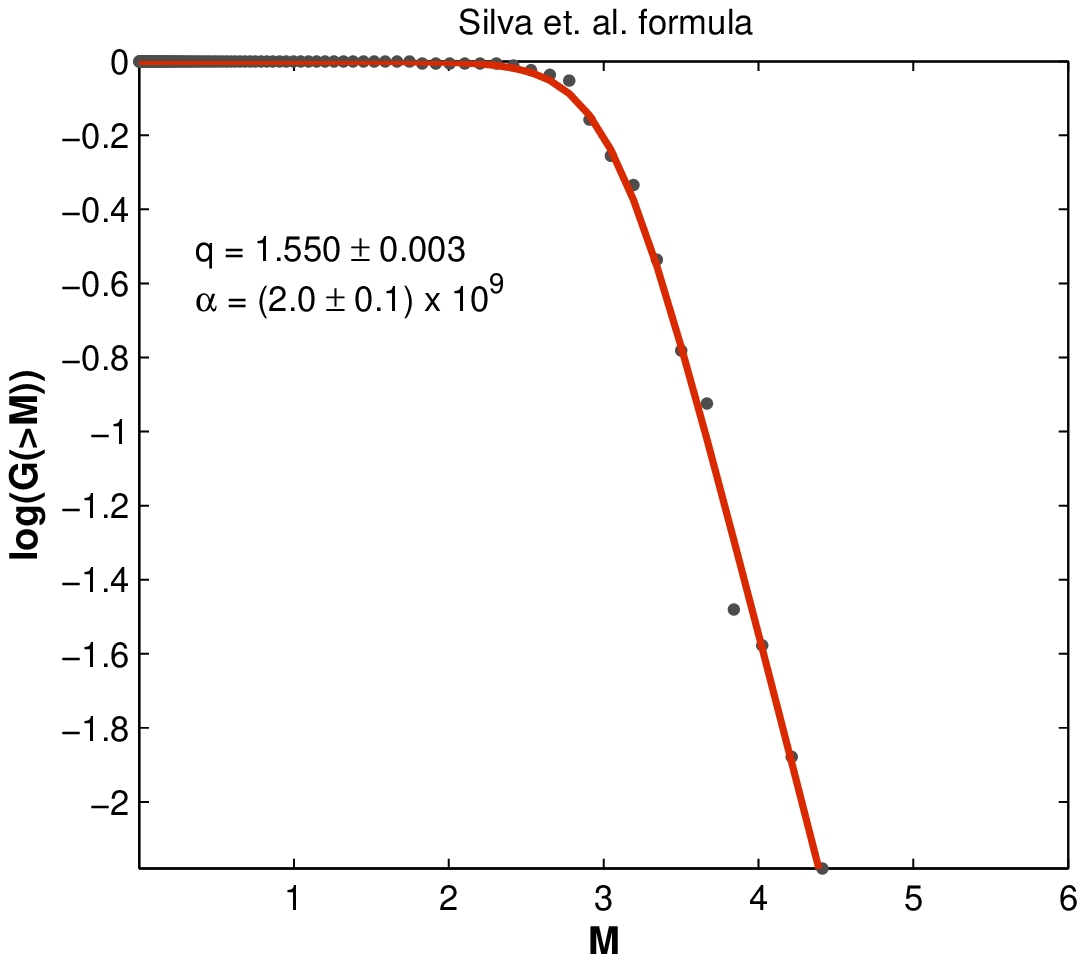}} 
     
	\subfloat[0-200km]{\label{subfig:q_ath3}\includegraphics[width=0.4\textwidth]{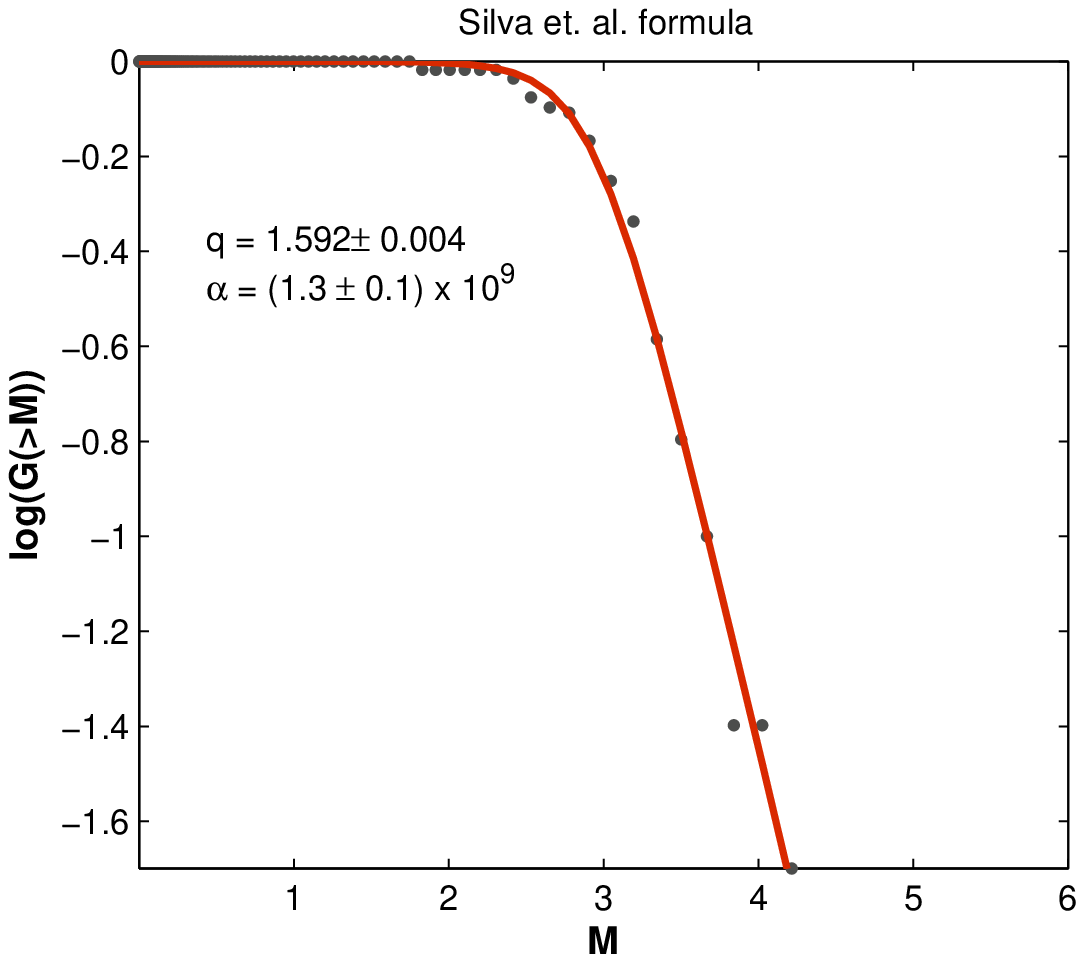}} 
	\subfloat[0-160km]{\label{subfig:q_ath4}\includegraphics[width=0.4\textwidth]{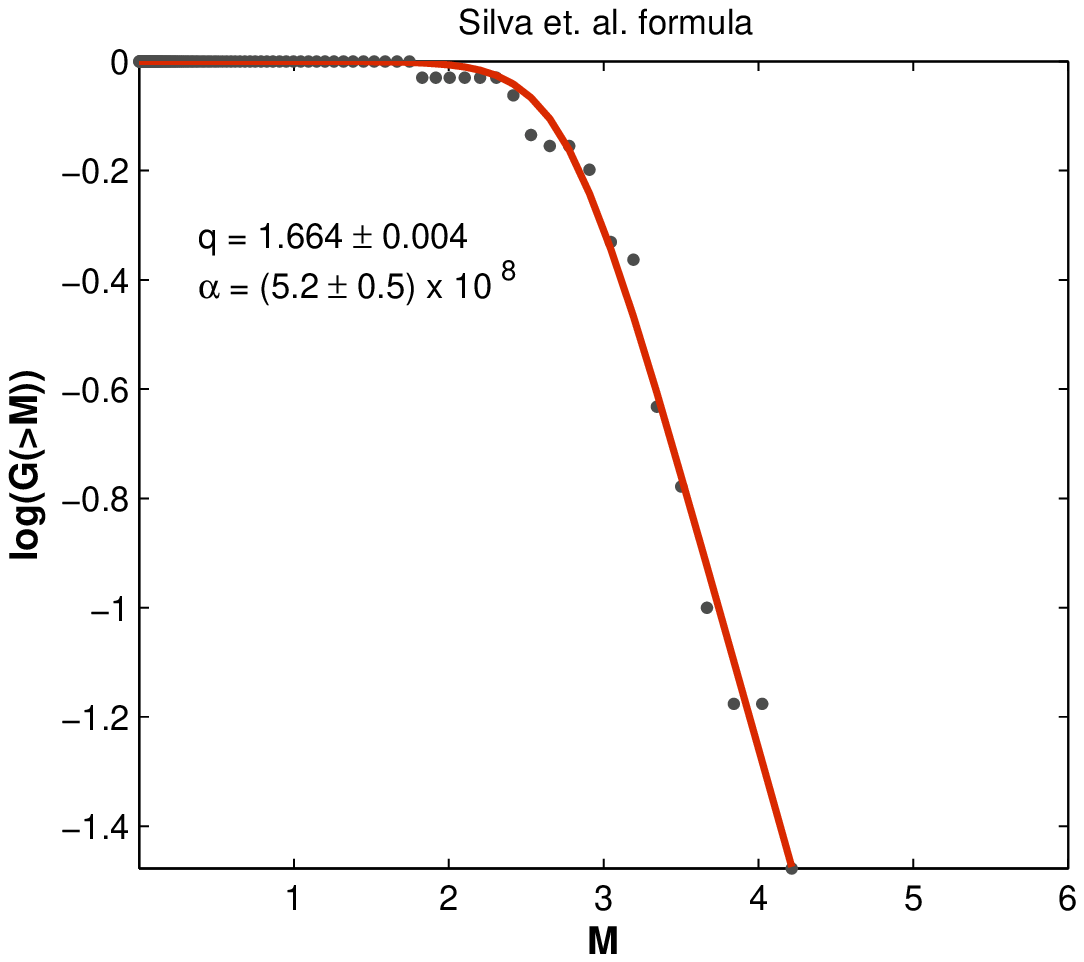}}      
       
	\caption{We use Eq. (\ref{eq:silva}) to fit the seismic data in terms of the relative cumulative number of EQs included in four different geographic areas around the Athens EQ epicenter.}
\label{fig:q_ath}
\end{figure}

Four different geographic areas were selected around the Athens EQ epicenter for the period under study: $0-400km$, $0-300km$, $0-200km$ and $0-160km$ respectively. In Fig. \ref{fig:q_ath}, we present the fitting of Eq. (\ref{eq:silva}) applied on the relative cumulative number of magnitudes ($G(>M)$) for the populations of EQs included in each selected geographic area. It is observed that the nonextensive $q$-parameter varies between $[1.532 \sim 1.664]$ with a relative small standard error ranging between $[0.003 \sim 0.004]$. 

\begin{figure}[h]
	\centering
	\subfloat[0-400km]{\label{subfig:qa_ath1}\includegraphics[width=0.4\textwidth]{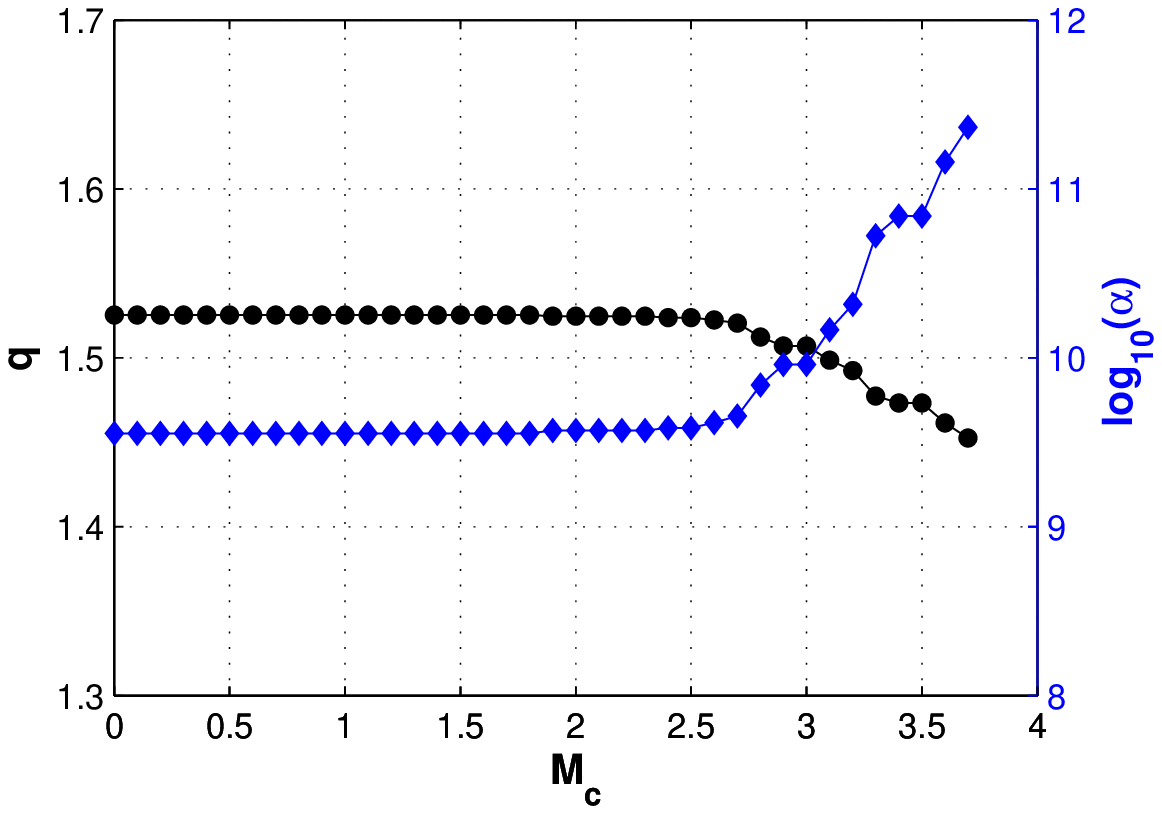}}             
	\subfloat[0-300km]{\label{subfig:qa_ath2}\includegraphics[width=0.4\textwidth]{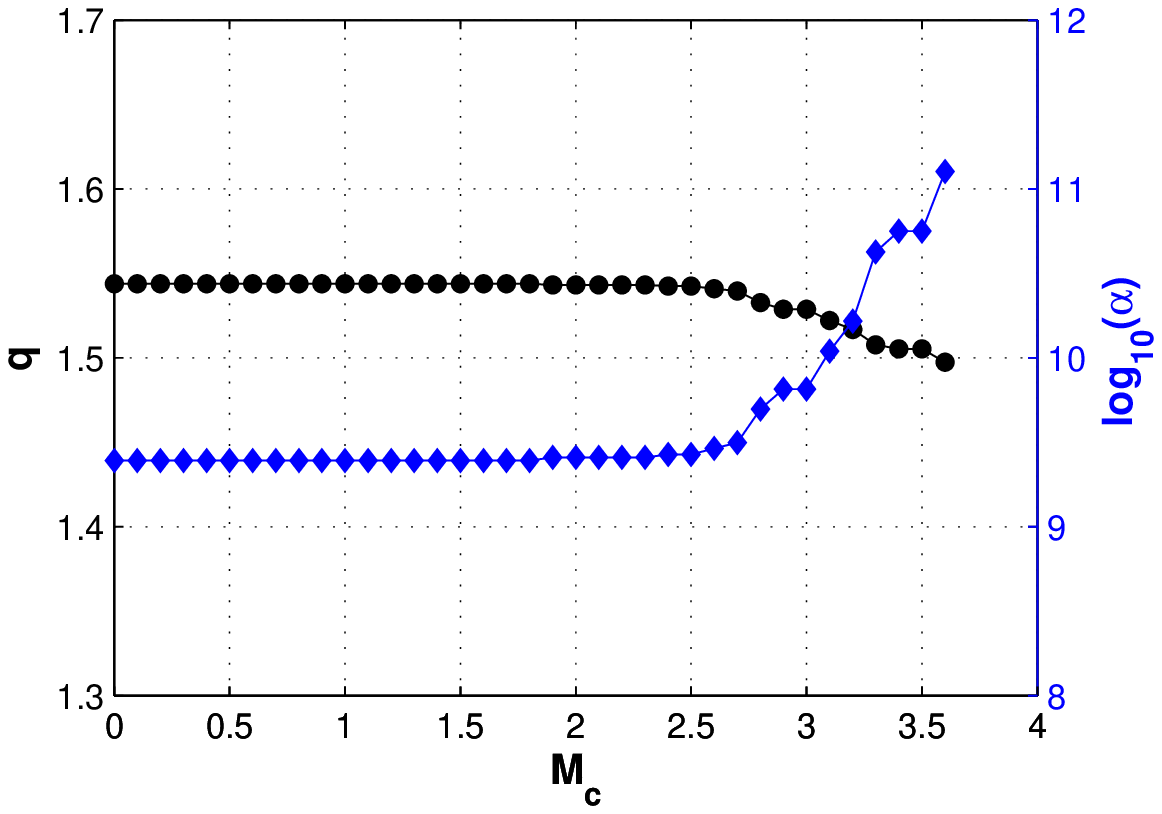}} 
     
	\subfloat[0-200km]{\label{subfig:qa_ath3}\includegraphics[width=0.4\textwidth]{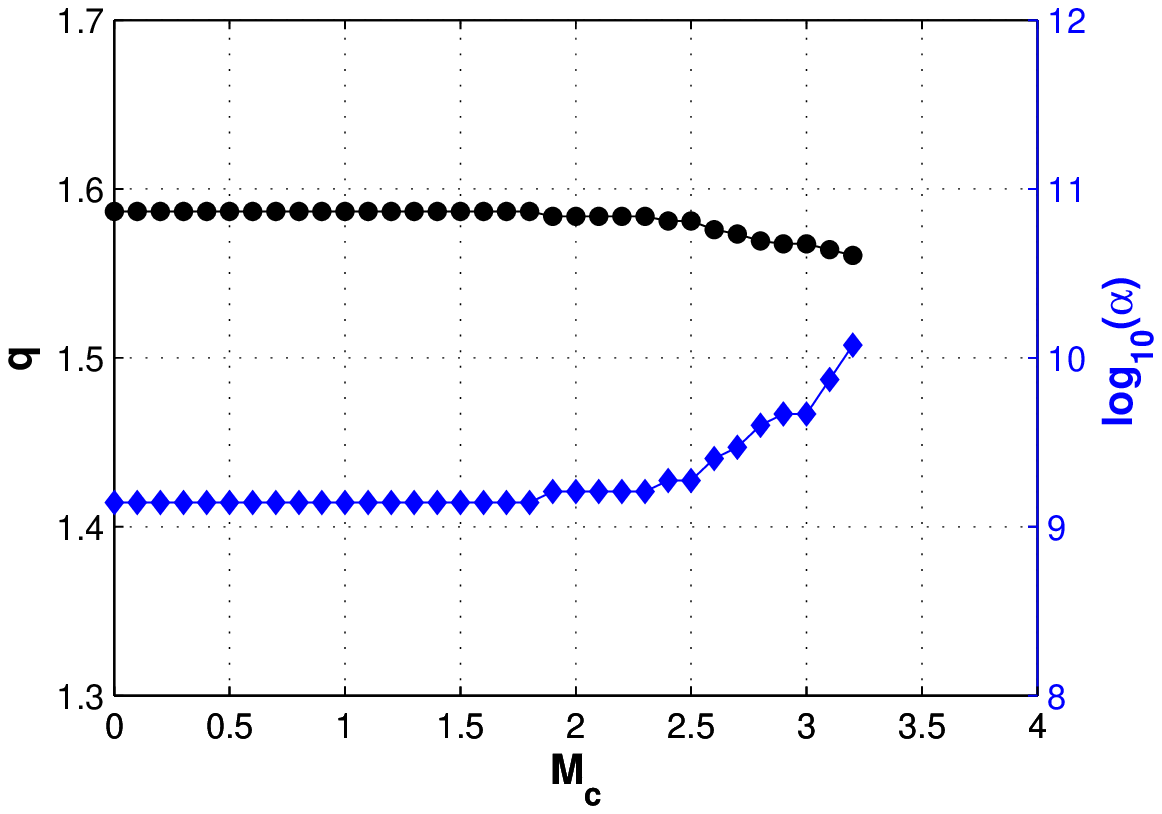}} 
	\subfloat[0-160km]{\label{subfig:qa_ath4}\includegraphics[width=0.4\textwidth]{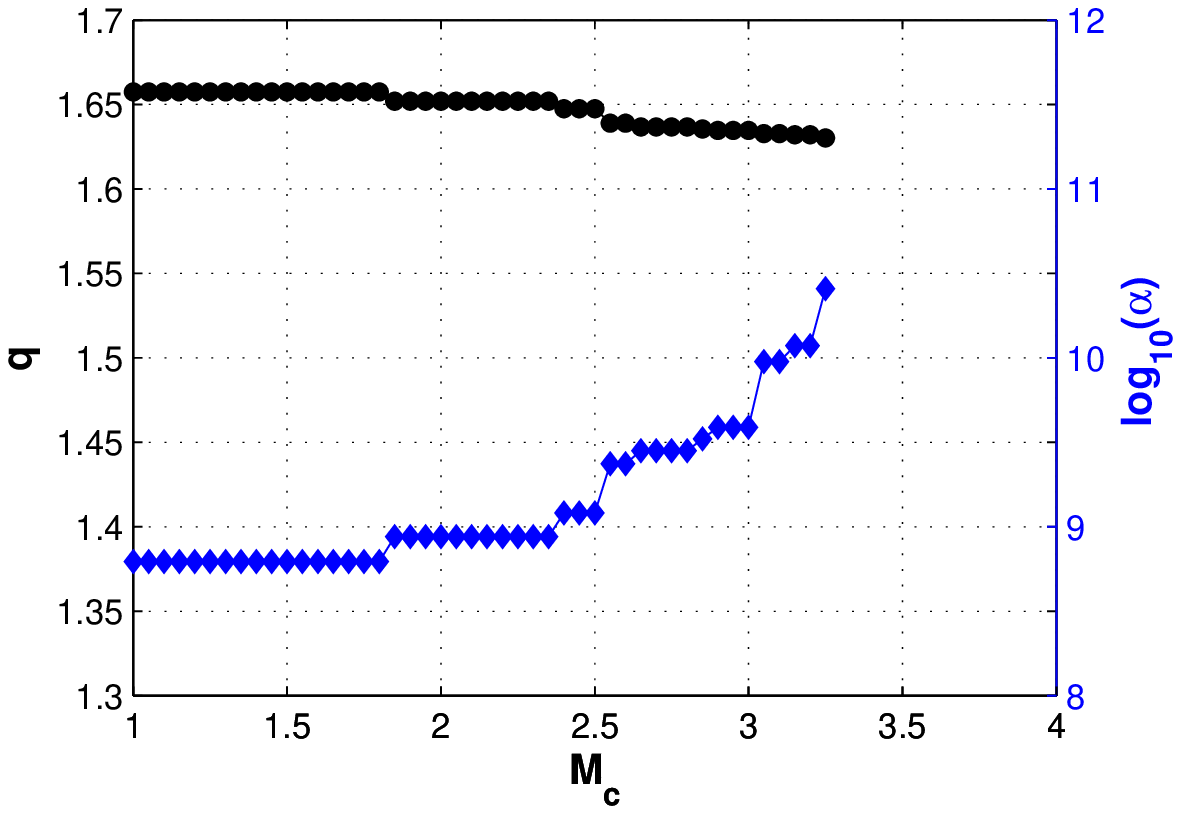}}      
       
	\caption{Variation of nonextensive parameter $q$ (see black-bullets) and the volumetric energy density $\alpha $ (see blue-rhombuses), for different thresholds of magnitudes of the detected EQs included in the period from 17-Aug-1999 00:01:39.80 up to 07-Sep-1999 01:56:49, for the range (a) 0-400km (b) 0-300km, (c) 0-200km and (d) 0-160km around the Athens EQ epicenter.}
\label{fig:qa_ath}
\end{figure}

Applying the same method as that described in Sec. \ref{sec:laquila}, in Fig. \ref{fig:qa_ath}, we present the variation of nonextensive parameter $q$ (see black-bullets) and the volumetric energy density $\alpha $ (see blue rhombuses), using different thresholds of magnitudes. It is observed that both the nonextensive $q$-parameter and the energy $\alpha$, present similar behaviour, for all the selected geographic areas around the EQ epicenter. More specifically, the nonextensive parameter $q$ (depicted with black bullets) remains relative stable with minor decrement at higher thresholds of magnitudes, while the characteristic value of the volumetric energy density $\alpha $ (depicted with blue rhombuses), increases at higher magnitude thresholds ($M_c$).

\subsection{Analysis of preseismic kHz EM emissions observed prior to Athens EQ}
\label{sec:ath_em}

\begin{figure}[h]
\begin{center}             
\subfloat[3 kHz NS]{\includegraphics[width=0.33\textwidth]{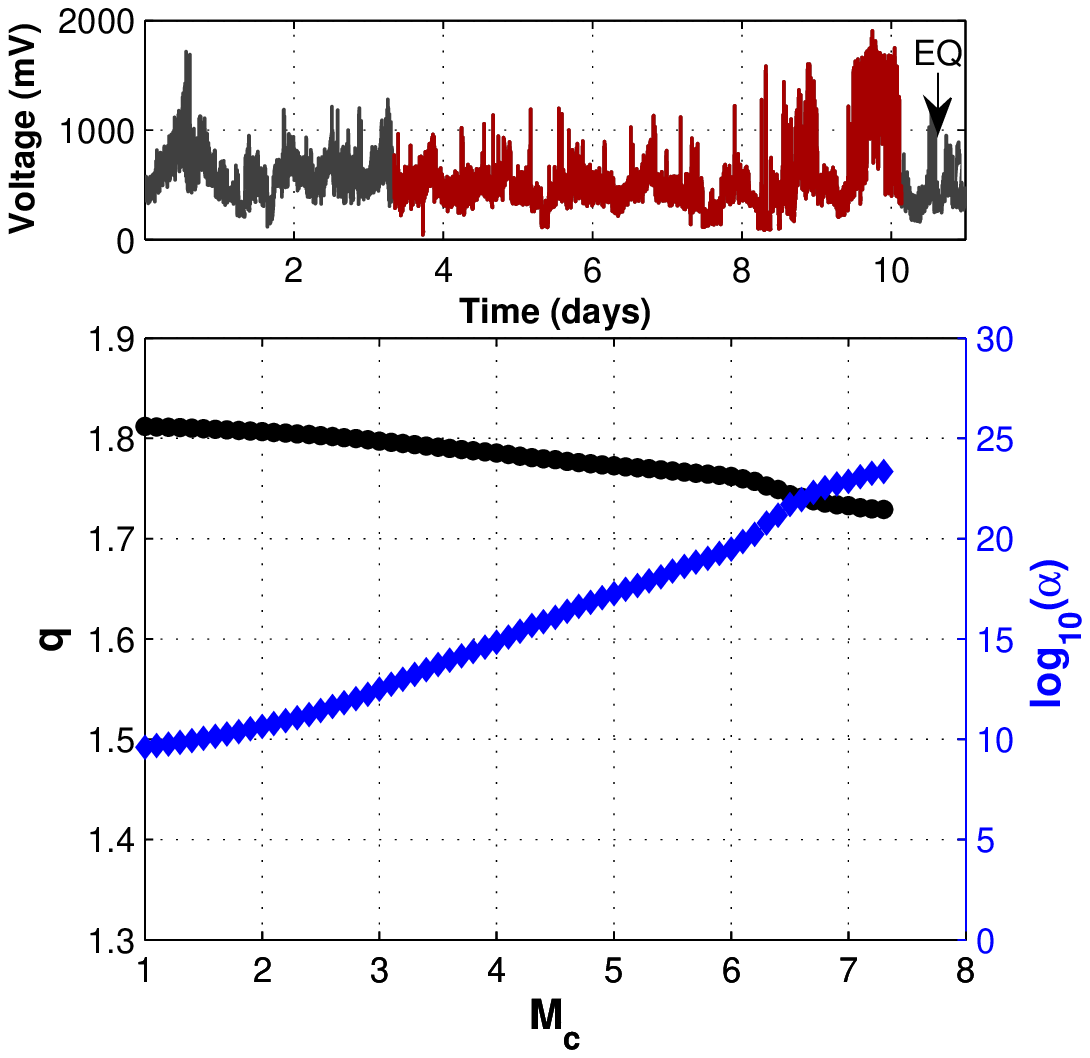}}
\subfloat[3 kHZ EW]{\includegraphics[width=0.33\textwidth]{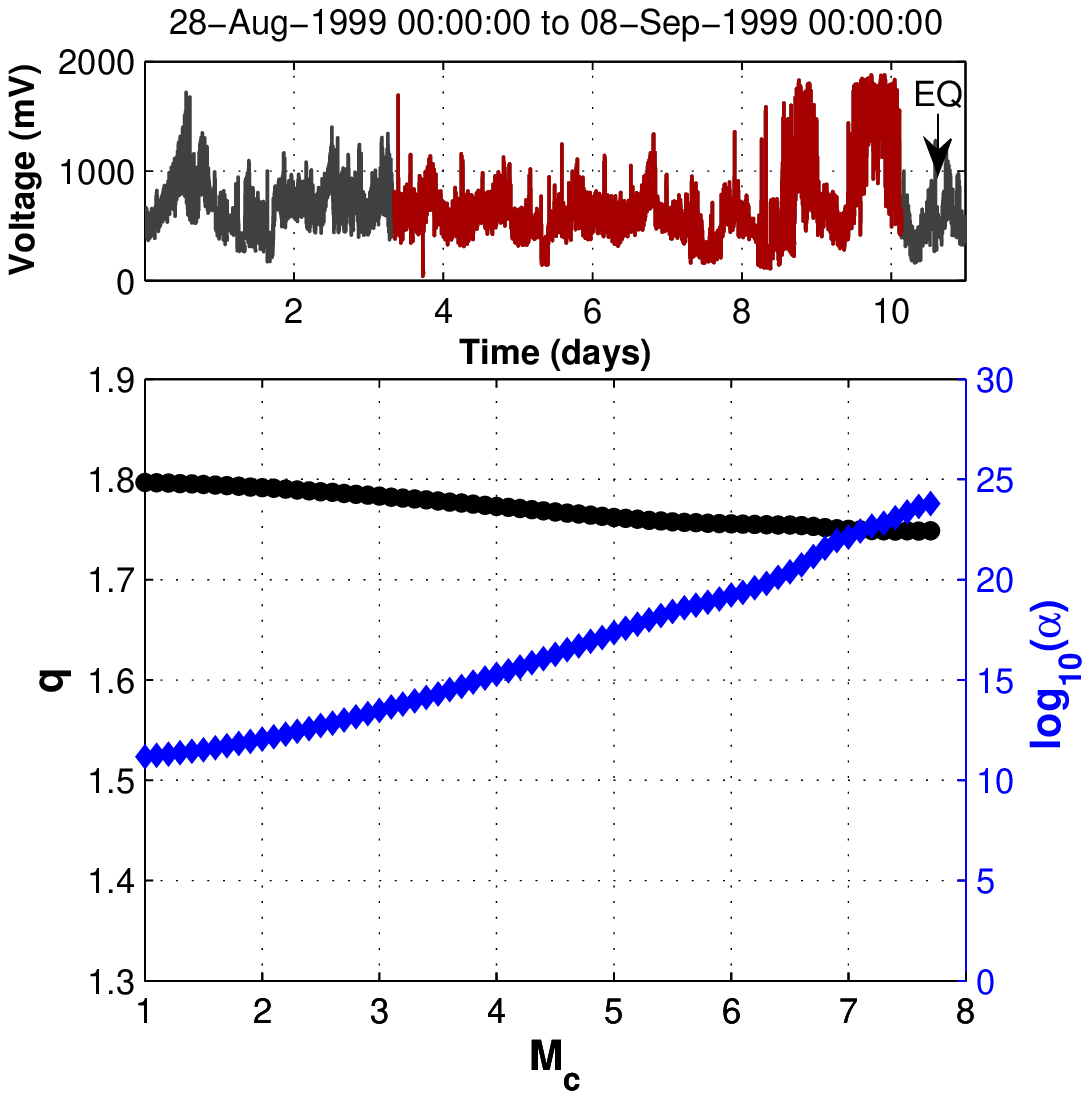}}
\subfloat[3 kHz V]{\includegraphics[width=0.33\textwidth]{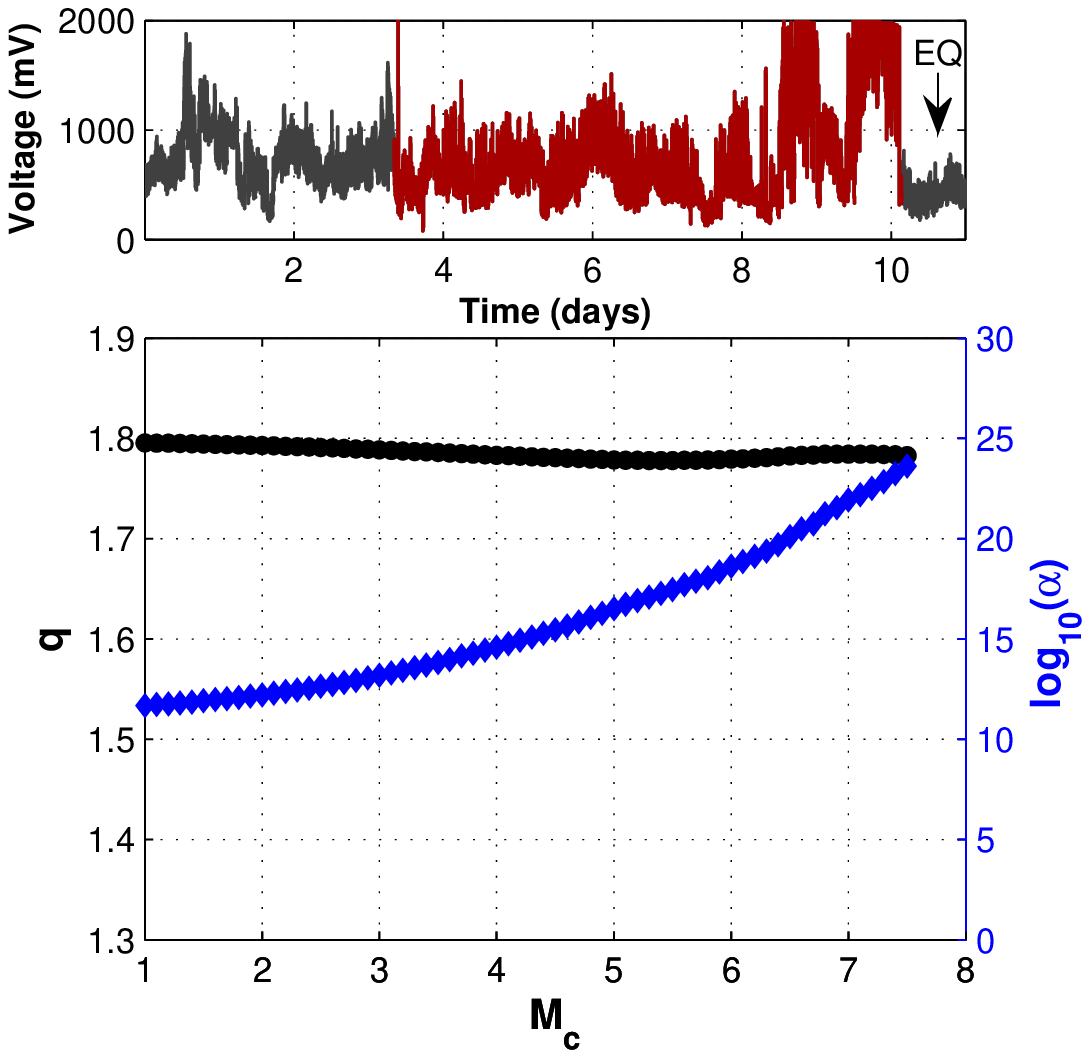}}

\subfloat[10 kHz NS]{\includegraphics[width=0.33\textwidth]{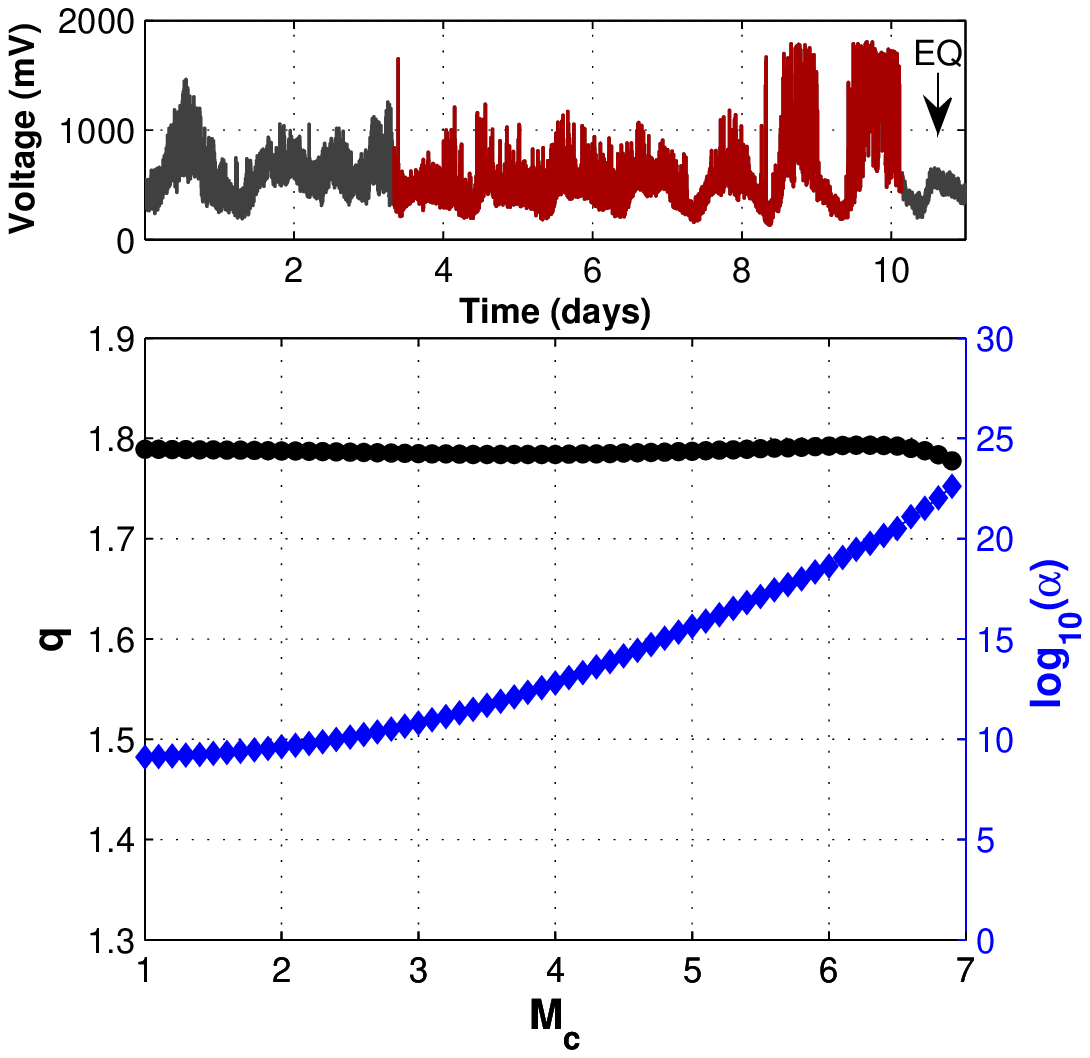}}
\subfloat[10 kHz EW]{\includegraphics[width=0.33\textwidth]{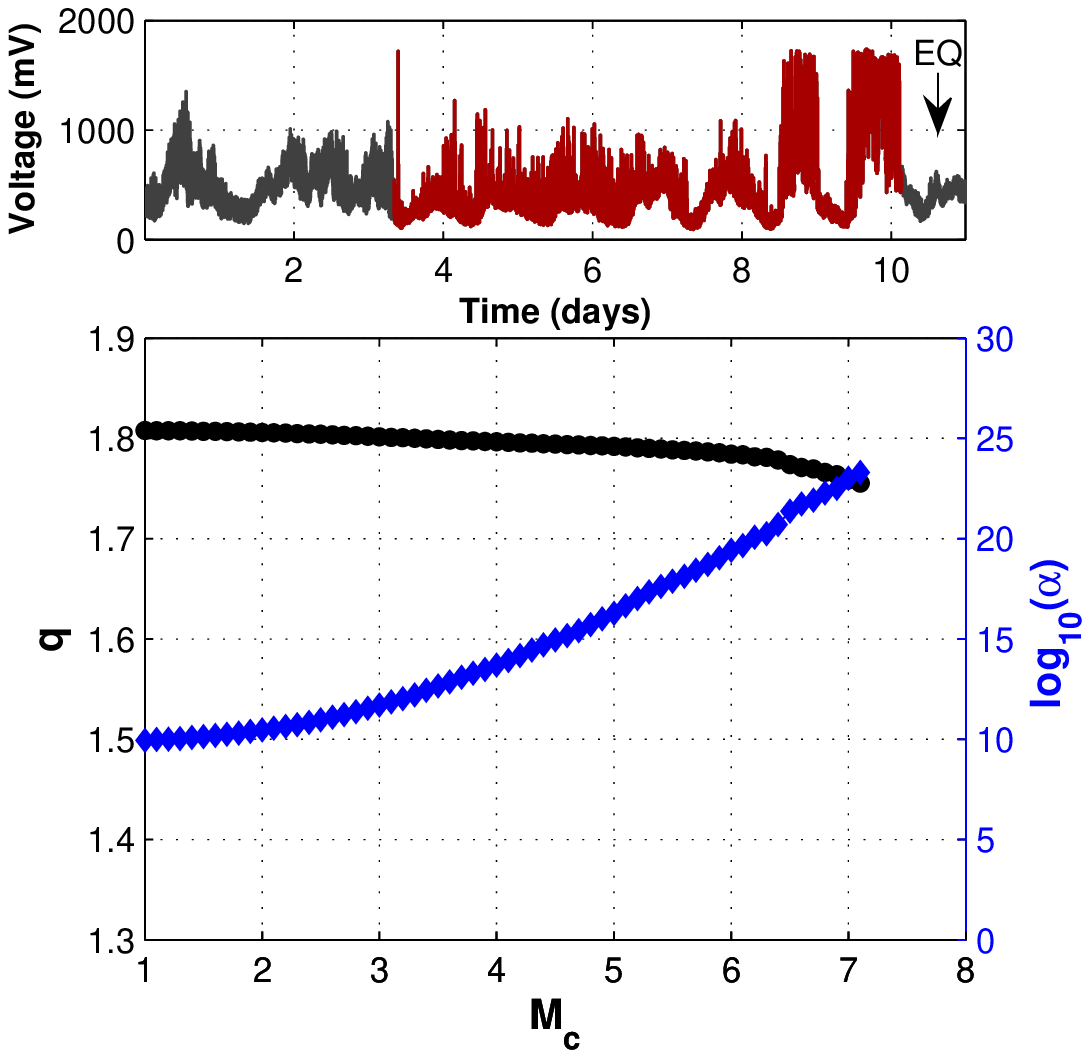}}
\subfloat[10 kHz V]{\includegraphics[width=0.33\textwidth]{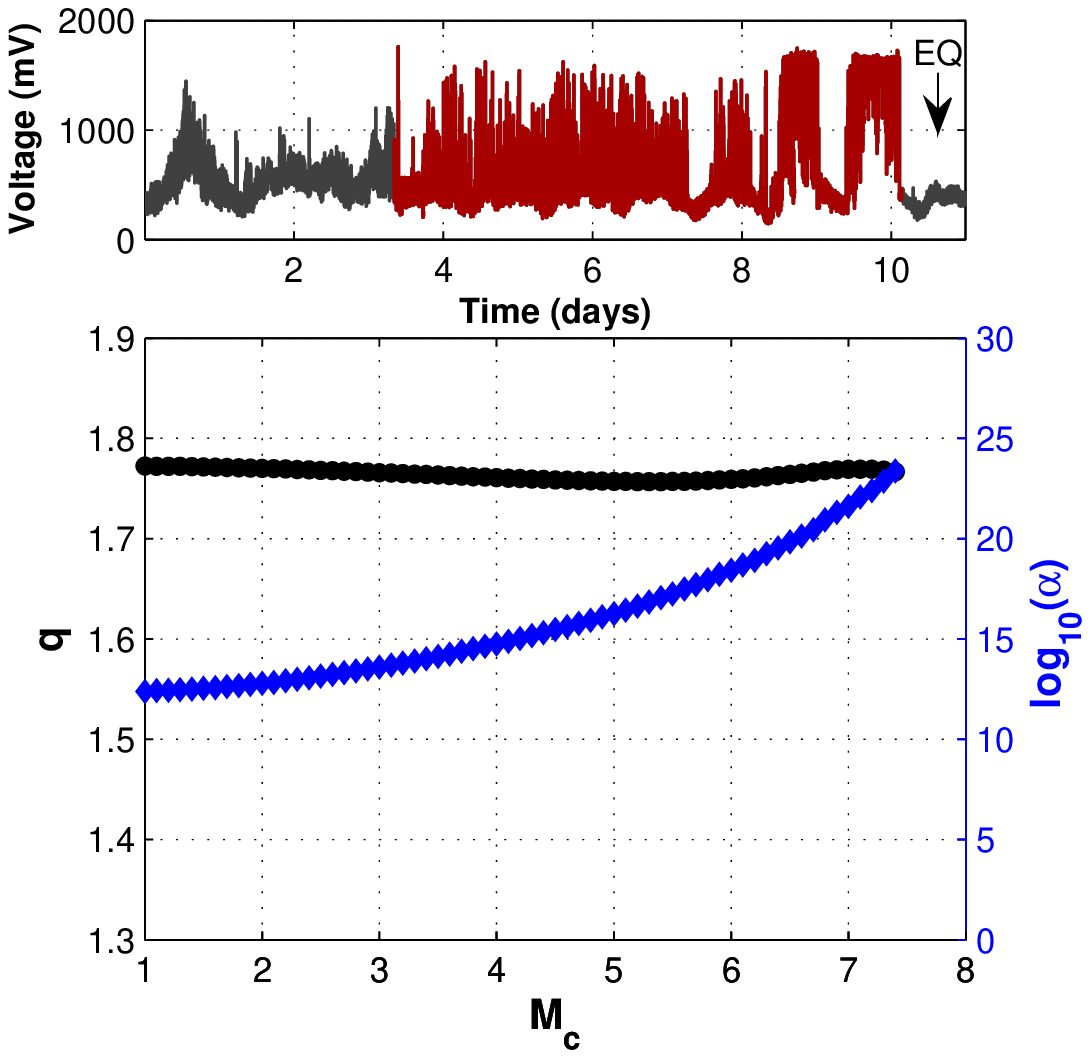}}

\end{center}
\caption{Six observed magnetic fields recorded on the period from 28-Aug-1999 00:00:00 to 08-Sep-1999 00:00:00. The selected red parts of each recorded signal refer to the period where the well justified for their seismogenic origin kHz EM anomalies have been observed (\citep{Kapiris2004,Contoyiannis2005,Contoyiannis2008,Papadimitriou2008}). The bottom charts of each sub-figure, refer to the corresponding variation of the nonextensive $q$-parameter (black-bullets) and the volumetric energy density $\alpha $ (blue-rhombuses), for different thresholds of magnitude cut-off ($M_c$).}
\label{fig:em_vars}
\end{figure}

The analysis here, is focused on the well documented \citep{Kapiris2004,Contoyiannis2005,Contoyiannis2008,Papadimitriou2008} kHz EM activity (Fig. \ref{fig:em_vars}, upper sub-charts) observed before the Athens EQ. The the six observed magnetic field strengths recorded by the 3 kHz (NS, EW and V) and 10 kHz (NS, EW and V) sensors, are analyzed, in terms of the nonextensive Eq. (\ref{eq:silva}). The top charts of each sub-figure depicted in Fig. \ref{fig:em_vars}, refer to the six observed magnetic fields recorded on the period from 28-Aug-1999 00:00:00 to 08-Sep-1999 00:00:00. Note that the Athens EQ occurred on 07-Sep-1999 07-Sep-1999 01:56:50 UTC as shown by the black-arrow. The selected red parts refer to the period where the kHz EM anomalies observed, which in turn have been well justified for their seismogenic origin from recent studies \citep{Kapiris2004,Contoyiannis2005,Contoyiannis2008,Papadimitriou2008}. 

We note that in a recent study \citep{Minadakis2012a} the six observed magnetic fields have been analyzed in the context of nonextensive Tsallis statistics. The analysis showed that the $q$-parameter lies within the interval  $q \in [1.78-1.82]$ with a relative small standard error ranging between $[0.0003 \sim 0.0006]$, for all the recorded channels. Herein, applying the same method as that described in Sec. \ref{sec:lac_em}, we further investigate the behaviour the nonextensive $q$-parameter and the volumetric energy density $\alpha $ from the perspective of self-affinity. More precisely, the bottom charts of each sub-figure contained in Fig. \ref{fig:em_vars}, depict the variation of the nonextensive $q$-parameter (black bullets) and the volumetric energy density $\alpha $ (blue rhombuses), for different thresholds of magnitudes ($M_c$). 

From Fig. \ref{fig:em_vars}, it is observed that the variation of the nonextensive $q$-parameter remains relative constant with minor decrement at higher thresholds of magnitudes. On the contrary, the energy density $\alpha $ mirrors the behaviour of $q$-parameter, presenting a relative increment at higher thresholds. These results evidently show the similarity with the results obtained in Sec. \ref{sec:ath_seis}, indicating the self-affine nature of fracture and faulting, from large-scale seismicity, to the fault-generation of a single EQ in terms of preseismic kHz EM emissions. In the next section we further provide further analysis and arguments that support the findings of this work. 

\section {Analysis of results - Discussion}
\label{sec:analysis}

Herein we recall that the main target of this work is to verify the self-affine nature of fracture and faulting in the framework of nonextensive Tsallis statistics; we examined whether it can be adequately explained by the nonextensive G-R type formula (Eq. \ref{eq:silva}). In this direction we note that the traditional G-R formula leads to the power-low distribution of magnitudes expressing the fractal nature of the system under study \citep{Sator2010,Turcotte1986,Sammis1986,Kaminski1998,Carpinteri2005}. On the other hand, Eq. (\ref{eq:silva}) is directly connected to the traditional G-R law (Eq. \ref{eq:b-value}), above some magnitude threshold through Eq. (\ref{eq:sarlis}) \citep{Sarlis2010,Telesca2010c,Telesca2012mle,Minadakis2011a,Minadakis2012a}. Moreover, since the released energy $\varepsilon$ is proportional to fragment size ($\varepsilon \propto r^3$) \citep{Silva2006}, it is reasonable to assume that the magnitudes of EQs which are rooted in the fracture of the population of fragments-asperities filling the space between fault planes, also follow a power-law distribution. 

The formula was mainly applied on both: 
\begin{enumerate}[(i)]
\item {the populations of EQs included in different radius around the epicenter of a significant seismic event (foreshock activity), and}
\item {the populations of EM-EQs which mirror the fracture of strong entities distributed along the main fault that sustain the system.}
\end{enumerate}

Indeed, analysis on both the populations of EQs included in foreshock activity and the populations of EM-EQs included in preseismic kHz EM emissions (see Figs. \ref{fig:q_lacuila},\ref{fig:em_vars_lac} and \ref{fig:q_ath}) has provided similar footprints of nonextensivity  verifying the aforementioned relation. More specifically, for the case of L'Aquila EQ the nonextensive $q$-parameter ranges between $[1.644 \sim 1.694]$ for the seismicity and between $[1.818 \sim 1.838]$ for the preseismic kHz EM emissions. For the case of Athens EQ, the nonextensive $q$-parameter ranges between $[1.532 \sim 1.664]$ for the seismicity and between $[1.78 \sim 1.82]$ for the preseismic kHz EM emissions. These calculated nonextensive $q$-parameters are in general agreement with those obtained from several independent studies, related to seismicities generated in various large geographic areas, involving the Tsallis nonextensive framework \citep{Silva2006,Telesca2010,Telesca2010b,Telesca2011,Matcharashvili2011}. We indicatively mention the study of Telesca \citep{Telesca2010b}, who found a value of ($q = 1.742$) by examining the preseismic activity of L'Aquila EQ, included in the period from 30-Mar-2009 to 06-Apr-2009. 

Focusing on the method for the estimation of the behaviour of parameters $q$ and $\alpha$, by using different magnitude thresholds ($M_c$), similar behaviour has also been found for both the populations of EQs included in foreshock activity and the populations of EM-EQs included in preseismic kHz EM emissions. Characteristically, from Figs. \ref{fig:qa_ath}, it is observed that the nonextensive parameter $q$ and energy $\alpha$, present similar behaviour, for all the selected areas around the Athens EQ epicenter. More specifically, it is observed that the nonextensive parameter $q$ (depicted with black bullets) remains relative stable for different magnitude thresholds. The situation changes for even larger thresholds of magnitude, where a relative decrement is observed. Such behaviour is also observed from the analysis of preseismic kHz EM emissions as shown in Figs \ref{fig:em_vars_lac} and  \ref{fig:em_vars}. This prospective decrement can be explained by the fact that the larger the magnitude threshold the larger the number of the omitted EQs. The absence of the small fractures along with the corresponding redistribution of stresses, contributes to the decrement of the correlation length during the fracture process \citep{Sornette2002}. The smaller magnitude threshold is the one that governs the overall system (e.q for $M_c=0$). It should also be noted that although the nonextensive parameter $q$ decreases at higher magnitude thresholds it still remains high verifying the strong correlations that have been developed. Concerning the volumetric energy density $\alpha $ (depicted with blue-rhombuses), its characteristic value, increases at higher thresholds of magnitude ($M_c$). Note that according to the fragment-asperity model (SCP), $\alpha $ is the coefficient of proportionality between fragment size and released energy \citep{Sotolongo2004,Silva2006}. This evidence is consistent with the hypothesis that larger EQs are rooted in larger and stronger entities \citep{Minadakis2011a,Papadimitriou2008} that sustain the system. 

\begin{figure}[h]
\begin{center}             
\includegraphics[width=0.7\textwidth]{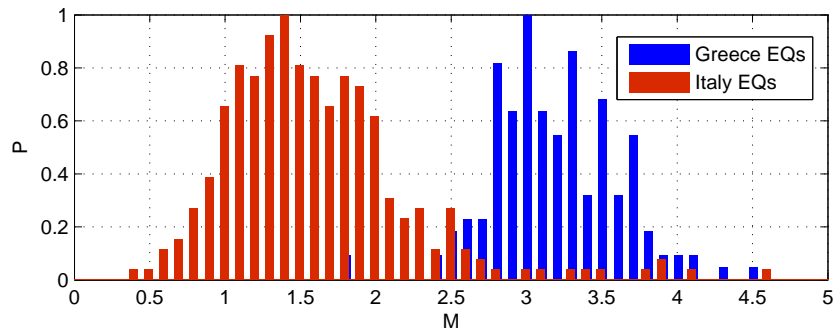}
\end{center}
\caption{Distribution of EQ magnitudes included in the area of 0-400km around the Italian EQ (red bars) and Greek EQ (blue bars), correspondingly.}
\label{fig:hists}
\end{figure}

A characteristic differentiation that should also be discussed here, mainly observed in the case of L'Aquila analysis of seismicity (see Fig. \ref{fig:qa_lacuila}), is the relative inclement of $q$ parameter at higher magnitude thresholds as opposed to the case of Athens EQ, where a relative decrement is observed (see Fig. \ref{fig:qa_ath}). This is also not a trivial result because if we get a closer look to Athens case in Fig. \ref{fig:qa_ath}, this variation is mainly observed for larger thresholds of magnitude in contrast to the case of L'Aquila EQ where the increment is observed at intermediate thresholds of magnitude. This evidence further enhances the suggestion that the small fractures along with the corresponding redistribution of stresses, contributes to the increment of the correlation length during the fracture process \citep{Sornette2002}. Indeed we calculated the mean magnitude of the EQs included in the area of 0-400km around the epicenter, for both the examined periods and has been found to be: $\bar{M} = 1.60$ for the case of Italy EQ and $\bar{M} = 3.00$ for the case of Greece. Furthermore, Fig \ref{fig:hists} shows the distribution of EQ magnitudes that correspond to the areas under study. It is clearly observed that in the case of the Italian territory, EQs with lower magnitudes are more frequent in contrast to the Greek territory. 

Along with these results, another feature that is also observed as approaching to the EQ epicenter, is that the smaller the area under study around the epicenter, the higher the nonextensive $q$-parameter. This evidence further verifies the increased correlations developed as approaching to the EQ epicenter. On the contrary the energy $\alpha $, presents a relative decrement that mirrors the behaviour of $q$-parameter. However, this mirroring behaviour between $q$ and $\alpha $, is still an open issue for the scientific community and also out of the scope of this study.

\section {Conclusions}
\label{sec:conclusions}

Building on the perspective of the self-affine nature of fracture and faulting theory, this work supports the hypothesis that the statistics of regional seismicity is a macroscopic reflection of the physical processes in the earthquake source. This suggestion implies that the activation of a single fault is a reduced self-affine image of regional seismicity. We used a recently introduced nonextensive Gutenberg \& Richter type formula which describes the EQ dynamics and includes two parameters: the entropic index $q$, which describes the deviation of Tsallis entropy from the standard Boltzmann-Gibbs entropy, and the physical quantity $\alpha $, which characterizes the energy density. Focusing on two cases of large EQs, we found similar behaviour of the corresponding variation of the parameters $q$ and $\alpha$, which are included in the nonextensive law for different thresholds of magnitudes applied on: (i) the population of EQs contained in different radii around the epicentre of a preceding large EQ and (ii) the population of ''fracto-electromagnetic earthquakes'' that are emerged during the fracture of strong entities distributed along the activated single fault sustaining the system. Analysis revealed that these two populations follow the same statistics, namely, the relative cumulative number of EQs against magnitude. Furthermore the results are further supported from recent studies in terms of the traditional Gutenberg-Richter law \citep{Sator2010,Turcotte1986,Sammis1986,Kaminski1998,Carpinteri2005}. Such analysis enhances the physical background of the underlying self-affinity.

\section{Acknowledgments}
The first author (G.M) would like to acknowledge research funding received from the Greek State Scholarships Foundation (IKY).

\end{document}